\documentclass[aip,rsi,reprint,graphicx]{revtex4-1} 
\usepackage{graphicx}
\usepackage{epsfig}
\usepackage{float}
\usepackage{color}
\usepackage{amsmath}
\usepackage[section]{placeins}
\usepackage{soul}

\draft 

\begin{document}


\title{Rigid platform for applying large tunable strains to mechanically delicate samples}

\author{Joonbum Park}
\thanks{These authors contributed equally.}
\email[]{jbpark0521@gmail.com}
\affiliation{Max Planck Institute for Chemical Physics of Solids, N\"{o}thnitzer Stra{\ss}e 40, 01187 Dresden, Germany}
\affiliation{Max Planck POSTECH Center for Complex Phase Materials, Pohang University of Science and Technology, Pohang 37673, Republic of Korea}

\author{Jack M. Bartlett}
\thanks{These authors contributed equally.}
\affiliation{Max Planck Institute for Chemical Physics of Solids, N\"{o}thnitzer Stra{\ss}e 40, 01187 Dresden, Germany}
\affiliation{Scottish Universities Physics Alliance (SUPA), School of Physics and Astronomy, University of St. Andrews,
St. Andrews KY16 9SS, United Kingdom}

\author{Hilary M. L. Noad}
\affiliation{Max Planck Institute for Chemical Physics of Solids, N\"{o}thnitzer Stra{\ss}e 40, 01187 Dresden, Germany}

\author{Alexander Stern}
\affiliation{Max Planck Institute for Chemical Physics of Solids, N\"{o}thnitzer Stra{\ss}e 40, 01187 Dresden, Germany}

\author{Mark E. Barber}
\affiliation{Max Planck Institute for Chemical Physics of Solids, N\"{o}thnitzer Stra{\ss}e 40, 01187 Dresden, Germany}

\author{Markus K\"{o}nig}
\affiliation{Max Planck Institute for Chemical Physics of Solids, N\"{o}thnitzer Stra{\ss}e 40, 01187 Dresden, Germany}

\author{Suguru Hosoi}
\affiliation{Department of Advanced Materials Science, University of Tokyo, Kashiwa, Chiba 277-8561, Japan}
\affiliation{Department of Materials Engineering Science, Osaka University, Toyonaka, Osaka 560-8531, Japan}

\author{Takasada Shibauchi}
\affiliation{Department of Advanced Materials Science, University of Tokyo, Kashiwa, Chiba 277-8561, Japan}

\author{Andrew P. Mackenzie}
\affiliation{Max Planck Institute for Chemical Physics of Solids, N\"{o}thnitzer Stra{\ss}e 40, 01187 Dresden, Germany}
\affiliation{Scottish Universities Physics Alliance (SUPA), School of Physics and Astronomy, University of St. Andrews,
St. Andrews KY16 9SS, United Kingdom}

\author{Alexander Steppke}
\email[]{steppke@cpfs.mpg.de}
\affiliation{Max Planck Institute for Chemical Physics of Solids, N\"{o}thnitzer Stra{\ss}e 40, 01187 Dresden, Germany}

\author{Clifford W. Hicks}
\email[]{hicks@cpfs.mpg.de}
\affiliation{Max Planck Institute for Chemical Physics of Solids, N\"{o}thnitzer Stra{\ss}e 40, 01187 Dresden, Germany}


\date{\today}

\begin{abstract}
Response to uniaxial stress has become a major probe of electronic materials. Tunable uniaxial stress may be applied using piezoelectric actuators, and so far two methods have been developed to couple
samples to actuators.  In one, actuators apply force along the length of a free, beam-like sample, allowing very large strains to be achieved. In the other, samples are affixed directly to
piezoelectric actuators, allowing study of mechanically delicate materials. Here, we describe an approach that merges the two: thin samples are affixed to a substrate, that is then pressurized
uniaxially using piezoelectric actuators. Using this approach, we demonstrate application of large elastic strains to mechanically delicate samples: the van der Waals-bonded material
FeSe, and a sample of CeAuSb$_2$ that was shaped with a focused ion beam.
\end{abstract}

\maketitle 


\section{Introduction}

Uniaxial stress has become a valuable probe of correlated electron systems. It is a qualitatively different probe from hydrostatic stress. For example, the critical temperature of the superconductor
Sr$_{2}$RuO$_{4}$ peaks strongly under uniaxial stress, while hydrostatic pressure causes a gradual decrease~\cite{Steppke17, Shirakawa97}. Uniaxial stress applied to YBa$_{2}$Cu$_{3}$O$_{6.67}$
suppresses superconductivity and stabilizes long-range charge modulation, while hydrostatic stress has the opposite effect~\cite{Souliou18,Kim18}. Strong nematic polarizability of Fe-based
superconductors has been revealed through application of anisotropic in-plane strain~\cite{Kuo16}. 

Recently-developed piezoelectric-based uniaxial pressure cells have allowed application of large uniaxial stresses at cryogenic temperatures. In Refs.~\onlinecite{Hicks14, Hicks14RSI, Barber19,
Steppke17, Kissikov18, Ikeda18, Mutch19}, the samples were prepared for these cells as free beams, whose ends were then affixed to the apparatus. The piezoelectric actuators apply strain to the sample
by applying displacement between the two ends. However, preparing samples as free beams is not appropriate for all materials and measurements. For preparing samples by hand, the minimum practical
sample length is $\sim$1~mm, and many potentially interesting materials are not available as single crystals even this large.  Moreover, a minimum mechanical strength is required to prepare samples as
free beams. When we attempted, for example, to prepare by hand a beam of the layered, van der Waals-bonded material FeSe, we found it all but impossible to avoid creasing the sample during handling.
Force applied to the beam deepened or flattened these creases instead of homogeneously straining the sample.

It has proved practical to strain small, mechanically delicate samples by affixing them directly to piezoelectric actuators~\cite{Chu12}. However in this case the sample strain is limited to that
which can achieved in the actuator, and if temperature is varied the unusual thermal contraction of piezoelectric actuators (they lengthen along their poling direction as they are cooled) may
introduce a large thermal strain. A further point of caution is that the surface of the actuator might not be uniform: the PICMA\textsuperscript{\textregistered} actuators from Physik Instrumente, for
example, have narrow slits for stress relief in the non-active surface layer. 

To merge the benefits of both approaches, we affix samples to a platform that is then mounted in uniaxial stress apparatus for application of large, tunable strains. Strain applied to the platform is
transmitted to the sample through the layer of epoxy between them. The idea is simple, and here we discuss practical engineering points involved in making it work. 

We also demonstrate that platforms can be used to apply strain to samples that have been microstructured with a focused ion beam (FIB). Microstructuring offers a number of possibilities, including
lower geometric uncertainty in measurement of transport coefficients, extreme aspect ratios for high-resolution measurements of resistivity~\cite{Moll16}, and measurements on very small samples. A
combination of ion beam milling and anisotropic strain, with the sample shaped for measurement of specific elastoresistivity coefficients, has been demonstrated in Ref.~\onlinecite{Rosenberg19}.

In section II below we discuss design of the platform, and of a uniaxial stress cell for pressurizing it. In section III, measurements of the strain actually achieved in the platform are presented. In
section IV, details of strain transmission from the sample to the platform are discussed, and in section V data on two samples are presented. One is a macroscopic sample of FeSe, a compound with
electronic nematic order whose transport properties are sensitive to lattice distortion, and the other is a microstructured sample of CeAuSb$_2$, a heavy-fermion compound with antiferromagnetic order
that is substantially altered by strain.

\section{Design} 

\begin{figure}[ht!b] 
\includegraphics[width=85mm]{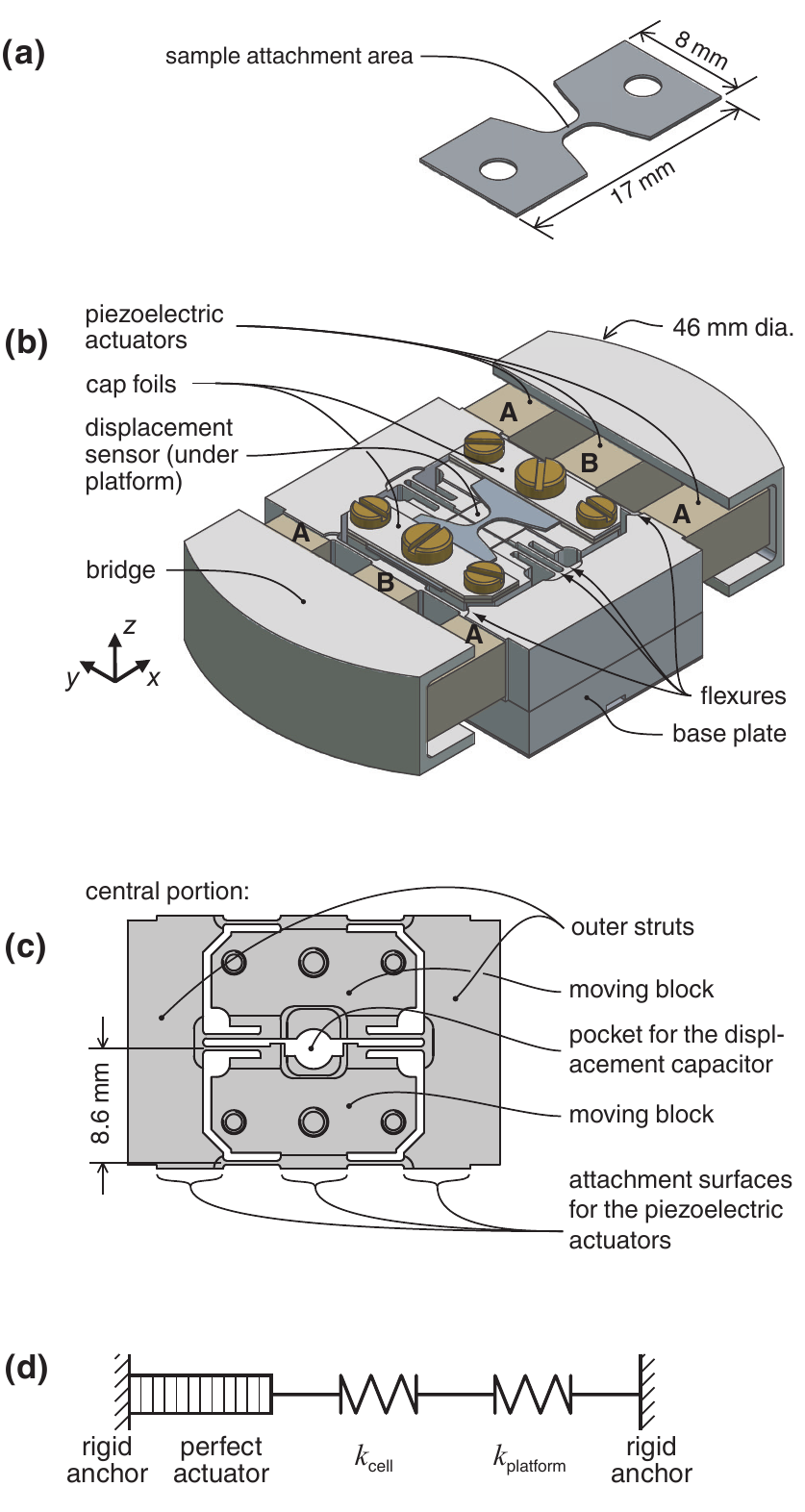}
\caption{\label{apparatusOverview}Illustration of the apparatus used here. \textbf{(a)} The platform, to which the sample gets affixed.  \textbf{(b)} The stress cell used in this work. The flexures
serve to allow longitudinal motion, while resisting torques and transverse forces. \textbf{(c)} Top view of the central portion of the cell. \textbf{(d)} To determine how much the platform is strained
when the actuators are operated, the system can be modelled as a perfect actuator (which generates a specified displacement irrespective of the resisting force) in series with a spring of spring
constant $k_\text{cell}$ representing the deformability of the cell, and one representing the deformability of the platform. Here, $k_\text{cell} = 12$~N/$\mu$m.}
\end{figure}

Schematics of a platform and the piezoelectric uniaxial stress cell used in this work are shown in
Fig.~\ref{apparatusOverview}(a,b). To understand expected performance, the specifications of the cell and
platform should be considered together. We discuss the cell first, then the platform, then the combined unit.

The cell is derived from the design presented in Ref.~\onlinecite{Hicks14RSI}, although in contrast to that cell the present cell has a symmetric configuration [Fig.~\ref{apparatusOverview}(b)]. The
central portion and bridges are made of titanium. The central portion [Fig.~\ref{apparatusOverview}(c)] consists of two outer struts connected by flexures to two inner moving blocks. These outer
struts are rigidly joined to a base plate and may be considered as fixed. The flexures serve to guide the moving blocks. They have a low spring constant against the intended longitudinal motion, but a
much higher spring constant against other motions.  Piezoelectric actuators on each side of the cell are used to apply displacement between the outer struts and moving blocks. For example, extension
of the actuators labelled A and contraction of those labelled B in Fig.~\ref{apparatusOverview}(b), through application of positive and negative voltages respectively, pulls the moving blocks outward
and tensions the platform.

To facilitate mounting of platforms the cell has a flat upper surface. The figure illustrates a mounting scheme in which the outer tabs of the platform are clamped under cap foils, a design intended
to allow rapid exchange of platforms while protecting them from torques applied while tightening the clamping screws.  Alternative mounting methods could also be devised. Placing the platform on top
of the cell means that the moving blocks experience torque about the $y$ axis when the cell is operated: the force applied by the actuators is not aligned with the resisting force from the platform.
The flexures resist this torque with a high spring constant.

As illustrated in Fig.~\ref{apparatusOverview}(d), the cell can be modelled as a perfect actuator (meaning an actuator that applies a specified displacement irrespective of the resisting force) in
series with a spring of spring constant $k_\text{cell}$ which represents the elastic compliance of the cell itself. If $k_\text{cell}$ is less than the spring constant of the platform, then the
displacement generated by the actuators goes mostly into deforming the cell itself, rather than the platform.  We present in the appendix an approximate calculation of $k_\text{cell}$, obtaining
$k_\text{cell} = 9$~N/$\mu$m. About half of this compliance comes from rotation of the moving blocks under the torque that they experience. In other words, $k_\text{cell}$ could be approximately
doubled by placing the sample / platform on the axis of the actuators. A compact, symmetric cell design in which the sample and actuators are aligned is presented in Ref.~\onlinecite{Kostylev19}; the
design here prioritizes a large mounting surface over maximum spring constant.

The spring constant of the cell was then measured at room temperature by applying a force using a spring of
known spring constant, and using a laser interferometer to measure the resulting displacement. The result is
in reasonable agreement with the calculation: 12~N/$\mu$m. Further details are given in the Appendix. 

We now discuss the platform design.  To achieve large strains, we introduce a short, narrow section in the middle to which the sample is mounted and in which applied force is concentrated, resulting
in a bowtie shape of the platform. The large tabs facilitate handling and mounting to the cell.  We fabricated platforms from two materials, 0.2 mm-thick temper annealed grade 2 titanium
foil~\footnote{Goodfellow, grade 2 (chemistry only) temper annealed titanium foils (Part No. TI000400), post-processed by KMLT Dermicut GmbH, Neukirch, Germany.} and 0.2 mm-thick fused quartz
plate~\footnote{Plan Optik AG, Eisoff, Germany, 4 inch-diameter, 200~$\mu$m-thick quartz wafers, post-processed by Lasercut24, Golmsdorf, Germany.}. We used titanium because its thermal contraction
matches that of the cell (though there may be small differences due to differing grain structure), and quartz because it is a thermally conductive insulator with a manageable Young's modulus: 73~GPa
for fused quartz at room temperature. (Sapphire, a more common choice when a thermally conductive, electrically insulating material is required, has a Young's modulus of 460~GPa, far higher than that
of either quartz or titanium.) Platforms of both materials were cut with a laser. 

\begin{figure}[htb]
\includegraphics[width=85mm]{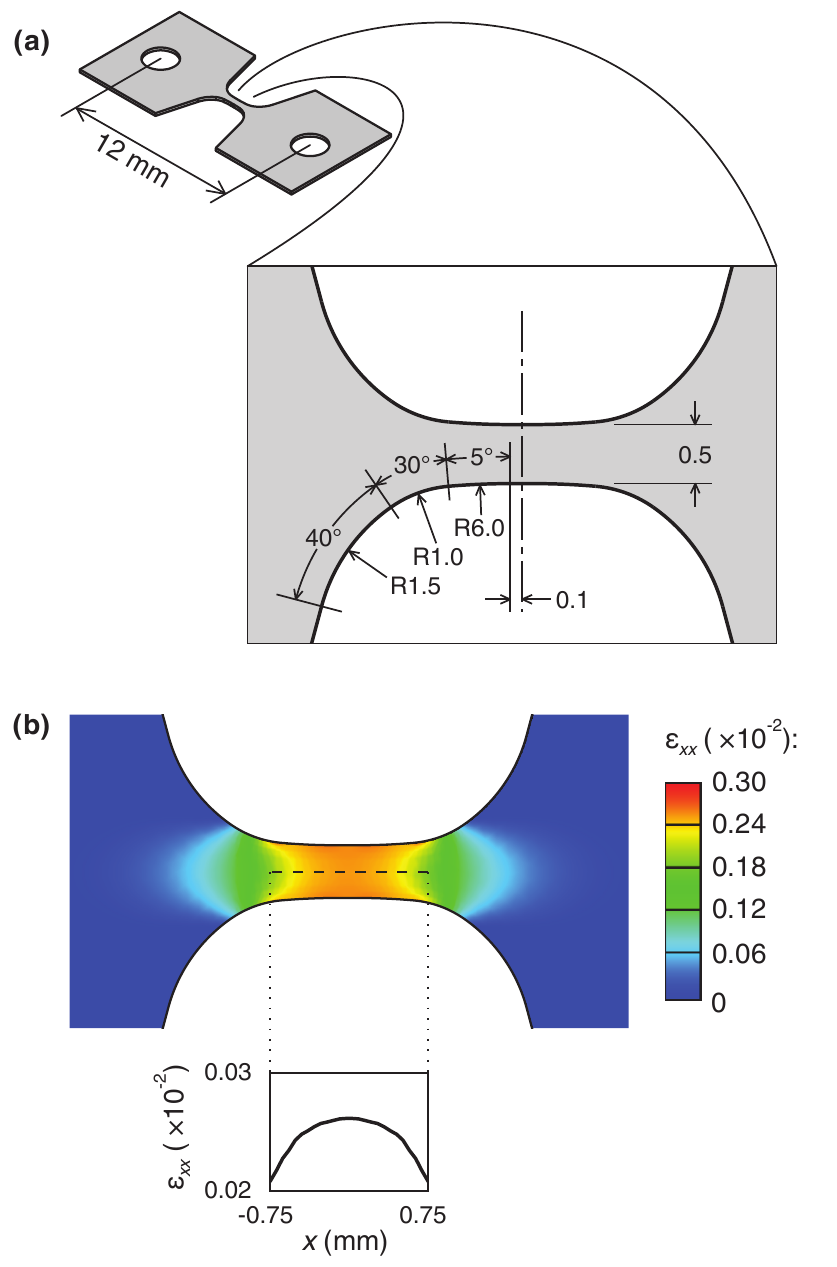}
\caption{\label{platformAnalysis}\textbf{(a)} The platform design used here; its thickness is 0.2 mm.
\textbf{(b)} Results of finite element analysis of the platform, using Autodesk Inventor\textsuperscript{\textregistered}, and taking titanium
for the platform material. In the simulation, a 10~$\mu$m displacement was applied between the mounting holes
in the platform. This simulation gives an effective length for the platform of 3.8~mm, a value that could vary
by $\sim$10\% depending on precisely which portions of the end tabs get locked to the cell. (In the simulation
shown here, the displacement was applied to the inner surfaces of the holes; in practice, it is the area
around the mounting holes that gets clamped.)}
\end{figure}

A key parameter for characterising platforms is their effective length $l_\text{eff}$, defined by $\varepsilon
= \Delta x / l_\text{eff}$, where $\Delta x$ is the displacement applied to the platform by the cell and
$\varepsilon$ is the longitudinal strain achieved in the neck of the platform. To a first approximation,
$l_\text{eff}$ is the length of the neck; however, it should be obtained through finite element analysis of
platform deformation. Our specific platform design is shown in Fig.~\ref{platformAnalysis}(a), and a
simulation of 10 $\mu$m displacement applied between the mounting holes [Fig.~\ref{platformAnalysis}(b)]
yields $l_\text{eff} = 3.8$~mm. In simulations $l_\text{eff}$ is found to vary by $\sim$10\% depending on
precisely which portions of the platform are taken to be locked to the cell, so it is not strictly a property
of the platform alone but of the cell and platform together. 

We now estimate the maximum strain achievable with this system, assuming elastic platform deformation. The spring constant of the platform is given by $E A / l_\text{eff}$, where $E$ is the Young's
modulus of the platform material and $A$ is the cross-sectional area of the neck. Taking $E = 103$~GPa for titanium gives $k_\text{platform} = 2.7$~N/$\mu$m, and $E = 73$~GPa for quartz yields
$k_\text{platform} = 1.9$~N/$\mu$m. At 1.5~K, the actuators can be operated safely at voltages between $-300$ and $+400$~V. At $-300$~V, the strain within the actuators is $\sim -7 \times 10^{-4}$,
and at $+400$~V, $\sim 8 \times 10^{-4}$, which yields a maximum displacement of $\sim$27~$\mu$m~\footnote{The actuators used in this cell are PICMA\textsuperscript{\textregistered} actuators from
Physik Instrumente, model number P885.10. Their response at 1.5~K is reported in Ref.~\cite{Barber19}.}. The fraction of this displacement that goes into the platform is
$k_\text{platform}^{-1}/(k_\text{cell}^{-1} + k_\text{platform}^{-1})$, which, taking $k_\text{cell} = 12$~N/$\mu$m, is 82\% for the titanium and 86\% for the quartz platforms. This yields, under an
assumption of elastic deformation, a maximum achievable strain of $5.8 \times 10^{-3}$ for the titanium platform, and $6.1 \times 10^{-3}$ for the quartz platforms. In reality the elastic limit of
grade 2 titanium is $\sim 2 \times 10^{-3}$, limiting the strain that can be achieved, and grade 5 titanium (Ti$_{0.90}$V$_{0.04}$Al$_{0.06}$) may be a better choice for high strains.

For electrical measurements on titanium platforms it is necessary to create an insulating layer between the
platform and the sample. We tested oxidation of the titanium surface by two methods: thermal and electrolytic.
For thermal oxidation, heating the platforms for four hours in air to 700$^\circ$ C resulted in an oxide layer
$\sim$1.6~$\mu$m thick.  The layer could be made thicker by heating for more time or at higher temperature;
however, it then flaked off more easily. We generally used electrolytic oxidation performed using a solution
of 10 g/L trisodium phosphate in water as an electrolyte. An applied voltage of 220~V for 15 minutes yielded
oxide films with a thickness of at least 200~nm. 

The electrolytic oxide layers were sufficiently robust to prevent electrical shorts between the platform and
samples placed by hand but were not highly reliable as insulation against evaporated gold contacts.
Furthermore, as noted in the introduction a major benefit of platforms is that they facilitate sample
preparation with a focused ion beam, however the ion beam quickly milled through the oxide layer and created
shorts through redeposited material. Therefore, a key advantage of quartz is that it is fully insulating. 

An advantage of a short $l_\text{eff}$ is that differential thermal contraction between the platform and
(titanium) strain cell can be compensated during temperature changes by operating the actuators, such that the
platform need not have a thermal contraction close to that of titanium. Fused quartz expands slightly during
cooling~\cite{Okaji95}; the differential thermal expansion between titanium and quartz upon cooling from 295
to 5~K is 0.16\%, corresponding here to a differential length change of $l_\text{eff} \times 0.16$\% =
6.0~$\mu$m. This is well within the range of the piezoelectric actuators of this cell.

\section{Experimental tests of the platform}

\begin{figure}[h!tb]
\includegraphics[width=85mm]{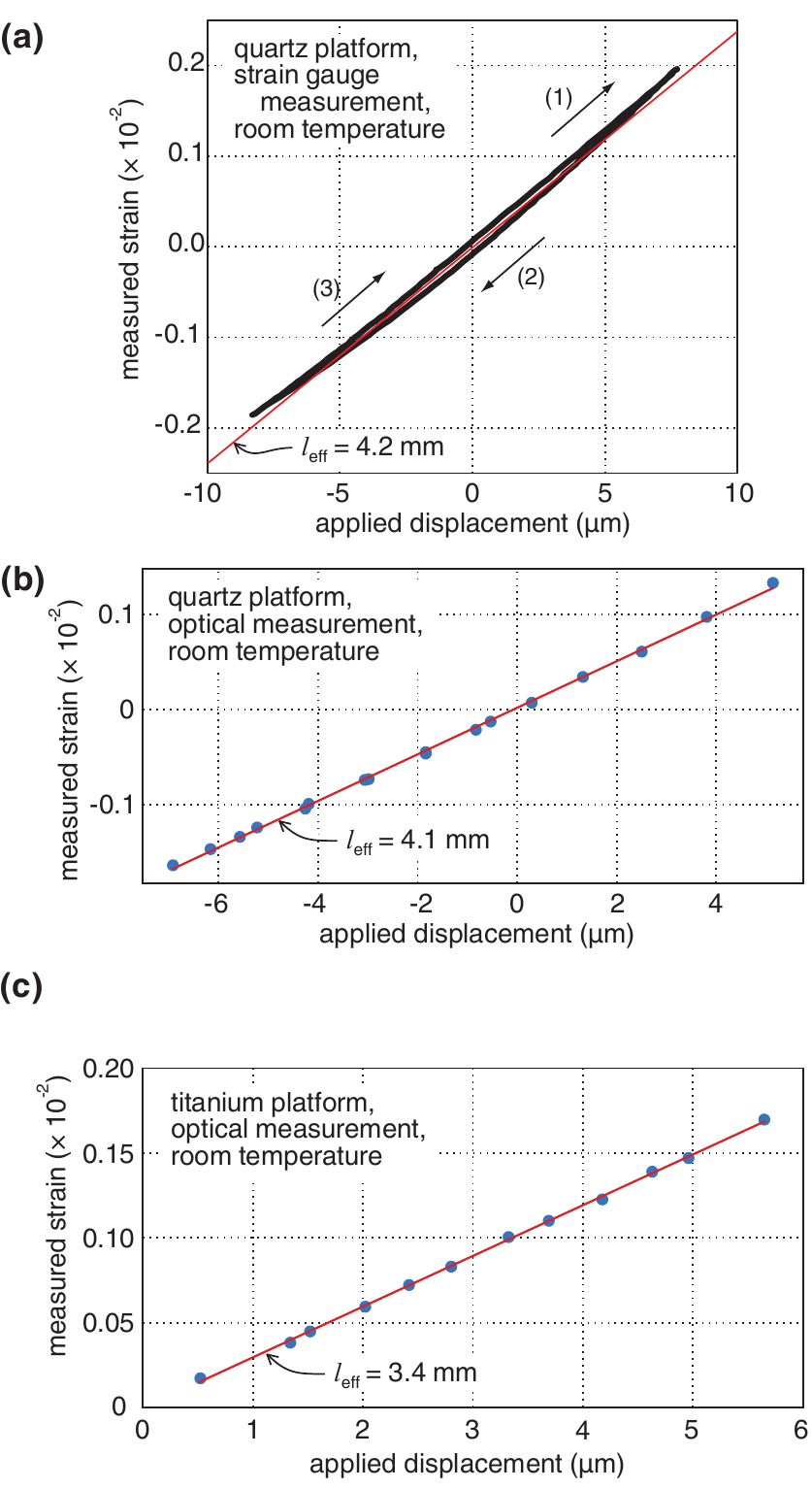}
\caption{\label{platformVerification}\textbf{(a)} Longitudinal strain achieved in a quartz platform, measured with a strain gauge, versus applied displacement, measured with the capacitive displacement sensor
built into the cell. The slope of the red line corresponds to $l_\text{eff} = 4.2$~mm. \textbf{(b)} Longitudinal strain achieved in a quartz platform, measured optically, versus applied displacement.
\textbf{(c)} Longitudinal strain achieved in a titanium platform, measured optically, versus applied displacement.}
\end{figure}

The strain actually achieved in the platform was tested by two means. In the first test, the strain achieved
in a quartz platform was measured at room temperature using a strain gauge affixed to the neck of the
platform, while the applied displacement was measured using the capacitive displacement sensor incorporated
into the cell. Results of the test using a strain gauge are shown in Fig.~\ref{platformVerification}(a).
Although there is minor hysteresis in the measured strain versus applied displacement, the effective length of
$l_\text{eff} = 4.2$~mm, obtained from a linear fit to the data, is close to the calculated effective length.
We note that this $l_\text{eff}$ is the empirical conversion constant between the displacement measured by the
sensor in the cell and the strain achieved in the platform. Due to torsional loading of the moving blocks the
actual displacement applied to the platform can differ by several percent from that measured by the sensor;
see the Appendix for details.

In the second test, the strain in a quartz platform was measured optically. A thin layer of silver epoxy was painted over the platform to create features whose positions could be tracked under a
microscope while displacement was applied to the platform. Pictures of the platform at different displacements were then analysed using image correlation \cite{Turner2015}. Results of the optical test
of the quartz platforms are shown in Fig.~\ref{platformVerification}(b). The effective length in this case was found to be $4.1$~mm, in good agreement with that found with the strain gauge. Finally, a
titanium platform was also tested optically at room temperature, keeping to strains below the elastic limit of the platform. Results are shown in Fig.~\ref{platformVerification}(c): $l_\text{eff}$ was
found to be 3.4~mm, slightly less than the calculated value.

\section{Calculations of strain transmission to the sample}

When using platforms, the sample will in general be thin, and the epoxy layer is likely to have much lower elastic moduli than the sample. The elastic compliance of the epoxy can limit strain
transmission to small samples.  When the sample and epoxy layer are both thin enough that the $z$ dependence of the strain within each can be neglected, and when the epoxy elastic moduli are low,
strain transmission from the platform to the sample can be characterized to a good approximation by a strain transmission length $\lambda$, a length scale over which the strain the sample adjusts to
match that in the platform. We note that this analysis will also apply to the thermal-expansion-based platforms reported in Refs.~\onlinecite{Sunko19, He18}, and also that it is not necessarily
desirable to make $\lambda$ as short as possible: increasing $\lambda$ reduces peak shear strains within the epoxy, potentially raising the maximum strain achievable in the sample before the epoxy
ruptures.

\begin{figure}[ptb]
\includegraphics[width=85mm]{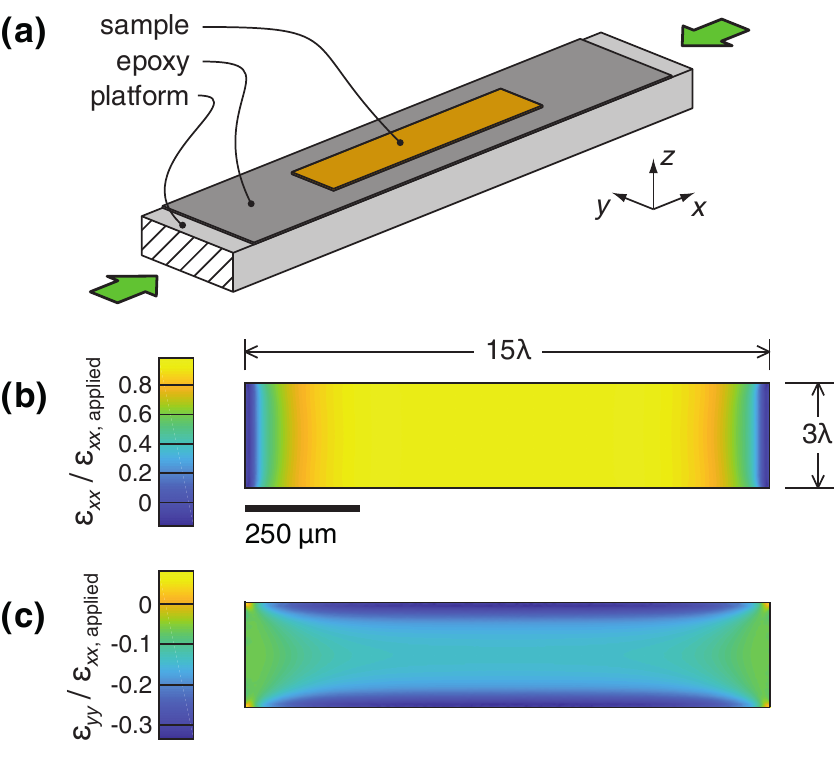}
\caption{\label{largeSampleCalculations}Simulation of strain transmission from the platform to the sample for a long sample. \textbf{(a)} Setup for the simulation. The epoxy is assigned a Young's
modulus of 4.5~GPa and the sample 100~GPa; further parameters are given in the text. \textbf{(b)} $\varepsilon_{xx}$ in the top surface of the sample. This sample is specified to have a length of
$15\lambda$ and width of $3\lambda$, where $\lambda$ is the strain transmission length, the length scale over which the strain in the sample adjusts to match that in the platform.  \textbf{(c)}
Transverse strain $\varepsilon_{yy}$ at the top surface of the sample. Because the sample width is not long compared with $\lambda$, $\varepsilon_{yy}$ is not uniform: at the sample edges it is set by
the Poisson's ratio of the sample, and in the center approaches that of the platform. To highlight the effect of Poisson's ratio mismatch, the platform has been assigned an unrealistic Poisson's ratio
of 0. The simulation was performed in {COMSOL}\textsuperscript{\textregistered}~\footnote{COMSOL Multiphysics v. 5.4. www.comsol.com. COMSOL AB, Stockholm, Sweden.}.}
\end{figure}

We consider a rectangular sample, as illustrated in Fig.~\ref{largeSampleCalculations}. We assume a sample length $l \gg \lambda$. In general, high strain homogeneity is achieved within the sample
when the width $w$ is either much less than or much greater than $\lambda$. In the former case, the transverse strain in the sample decouples from that in the platform, and is set instead by the
longitudinal strain multipled by the sample's Poisson's ratio. In the latter case, the transverse strain locks to that of the platform, which is the longitudinal strain multiplied by the platform's
Poisson's ratio. The strain transmission length was derived in Ref.~\onlinecite{Hicks14RSI}: $\lambda = \sqrt{Ctd/G}$, where $C$ is the relevant elastic modulus of the sample, $t$ the sample
thickness, $d$ the epoxy thickness, and $G$ the shear modulus of the epoxy.  For most epoxies $G$ is a strongly temperature-dependent parameter. At cryogenic temperatures, Stycast 1266 has a Young's
modulus of 4.5~GPa~\cite{Hashimoto80}, and taking a Poisson's ratio of 0.3 yields $G = 1.7$~GPa. If $C = 100$~GPa (a typical value for a metal) and $t = d = 10$~$\mu$m, then $\lambda$ comes to
76~$\mu$m.

In the narrow-sample limit, the $y$- and $z$-axis stresses in the sample are both zero, and $C$ is the Young's modulus of the sample. In the wide-sample limit, the transverse strain is fixed while the
$z$-axis stress is zero, and $C = C_{11} - C_{13}^2/C_{33}$, where $C_{ij}$ are components of the elastic tensor. For typical materials, these moduli are not drastically different, and the sample and
platform Poisson's ratios will also not differ drastically, and so whether the sample is in the narrow or wide limit is not highly important. 

FeSe, on the other hand, has a tetragonal-to-orthorhombic structural transition at $T_\text{s} = 90$~K,
and the distinction is important. In the vicinity of this transition its Young's modulus, for strains along the principal axes of the  distortion, is extremely small. The lattice however still resists
changes in unit cell area, and so $C_{11} - C_{13}^2/C_{33}$ remains substantial, at $\sim$40~GPa~\cite{Millican09, Zvyagina13}. Finite element simulation may be necessary to understand fully the
strain achieved in samples such as FeSe that have unusual elastic properties. The point of the discussion here is not to precisely map the strain in a sample, but to provide guidelines for setting
sample dimensions.

In Fig.~\ref{largeSampleCalculations}(b-c), results are shown of finite element simulation of the strain in a rectangular sample. The epoxy was assigned a Young's modulus of 4.5~GPa and an isotropic
Poisson's ratio of 0.3. The sample was assigned a Young's modulus of 100~GPa and isotropic Poisson's ratio of 0.3. The epoxy and sample thickness are both set to 10~$\mu$m. The sample length and width
are set to $15\lambda$ and $3\lambda$, respectively, \textit{i.e.} 1140 $\times$ 228~$\mu$m: we choose an intermediate width to highlight the effect of incomplete transmission of transverse strain.
The epoxy layer is assumed to have uniform thickness even across the sample edge; in reality the epoxy will wick up the sides of the sample, however the low elastic moduli of the epoxy means that the
effect of this on the strain in the sample will be minimal. The platform's neck has a cross section of $500 \times 200$~$\mu$m, and we assign a Young's modulus of 125~GPa. For the purposes of simulation the
platform is taken to have a constant cross section, and strain is applied to the platform by applying force to its end faces. The platform is assigned a Poisson's ratio of zero, an unrealistically low
value that is chosen to bring out in the simulation the effect of Poisson's ratio mismatch between the sample and platform.

Fig.~\ref{largeSampleCalculations}(b) shows the longitudinal strain $\varepsilon_{xx}$ at the upper surface of the sample. It is essentially zero at the sample ends, and then on moving towards the
center of the sample increases following a saturating exponential. Because the sample is long compared with $\lambda$, the strain in the center nearly matches that applied to the platform.

Fig.~\ref{largeSampleCalculations}(c) shows the transverse strain $\varepsilon_{yy}$. Along the edges, $\varepsilon_{yy}$ is controlled by the Poisson's ratio of the sample, whereas towards the center
it is controlled more by that of the platform. Because the width of this sample is neither long nor short compared with $\lambda$, and the platform and sample Poisson's ratios were chosen to be very
different, $\varepsilon_{yy}$ has low uniformity. 

\begin{figure}[ptb]
\includegraphics[width=85mm]{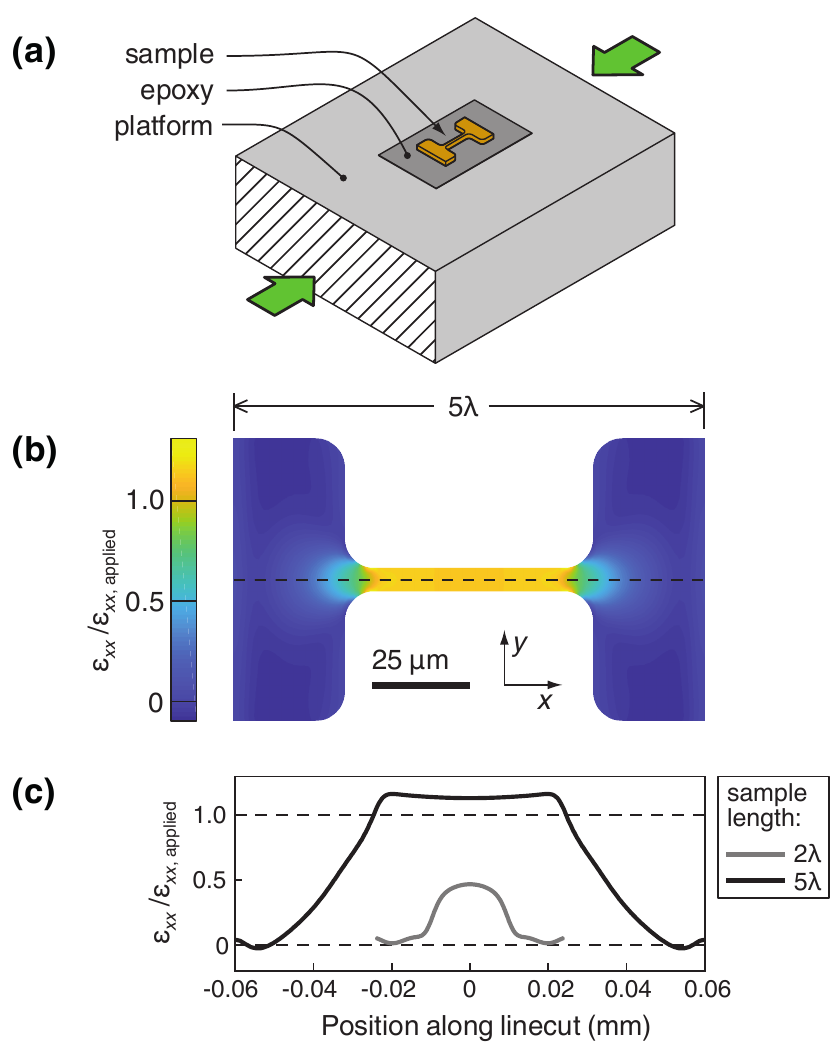}
\caption{\label{smallSampleCalculations}Simulation of strain transmission from the platform to the sample for a small sample. To achieve good strain transmission even when the sample is not long compared with $\lambda$, the sample
has been shaped, \textit{e.g.} through focused ion beam milling, into a narrow neck between two anchor tabs. \textbf{(a)} Setup for the simulation. Parameters are given in the text; the total sample
length is set to $5\lambda$. \textbf{(b)} $\varepsilon_{xx}$ in the top surface of the sample. The tabs are narrow compared with $\lambda$, and so are essentially unstrained, while the neck is highly
strained. \textbf{(c)} Line cut through the illustration in panel (b), and also a line cut for an even smaller sample. The thickness of this smaller sample was the same as for the larger sample,
10~$\mu$m, and its other dimensions were scaled to a total sample length of $2 \lambda$. Even with the shaping, this is too short for effective strain transmission. This simulation was performed in
{COMSOL}\textsuperscript{\textregistered}.}
\end{figure}

Samples that cannot be made long with respect to $\lambda$ can be shaped with FIB milling to achieve good strain transmission.  We illustrate the concept in Fig.~\ref{smallSampleCalculations}(a): the
center of the sample is milled into a narrow neck, and wide end tabs anchor the ends of this neck to the platform. The measurement would then be configured, for example in the placement of voltage
contacts, to measure the properties of the neck. For this simulation we set the epoxy thickness to 1~$\mu$m, a thickness that we have found to be achievable for smaller samples, and leave all other
parameters unchanged from the preceding simulation. 

The calculated profile of $\varepsilon_{xx}$ is shown in Fig.~\ref{smallSampleCalculations}(b).  The end tabs are essentially unstrained, because they are short along $x$ compared with $\lambda$,
however their area is sufficient that they couple to the platform and transfer substantial force from the platform to the neck. In fact, because the tabs themselves resist straining, the strain in the
neck overshoots that in the platform. 

In Fig.~\ref{smallSampleCalculations}(c) we also show results for an even smaller sample: still of thickness 10~$\mu$m, however with other dimensions scaled so that its total length is $2\lambda$. The
strain in the neck now considerably undershoots that in the platform; in other words even with this shaping this sample is too small for effective strain transmission. In general, shaping the sample
as presented here is a method to transfer strain effectively into smaller samples, however uncertainty in the thickness of the epoxy layer and in the epoxy elastic moduli will introduce uncertainty
into the strain actually achieved.

\section{Measurements of samples}

We first present results on FeSe, and then a microstructured sample of CeAuSb$_2$. FeSe has electronic nematic order, a spontaneous anisotropy in the electronic structure, below 90~K. In the vicinity
of this nematic transition, a high susceptibility towards electronic orthorhombicity causes the resistivity to respond very sensitively to lattice distortion~\cite{Hosoi16, Watson15}. Its structural
simplicity make it an appealing target for study, however it is a layered compound with van der Waals interlayer bonding, which makes samples mechanically delicate and difficult to strain. CeAuSb$_2$,
on the other hand, is mechanically more robust. It has an antiferromagnetic transition at 6.5~K, that is strongly altered under orthorhombic lattice distortion~\cite{Park18}. Its resistivity changes
strongly across this transition, providing an easy-to-measure signal that makes CeAuSb$_2$ a good test subject.

\begin{figure}[tbp]
\includegraphics[width=85mm]{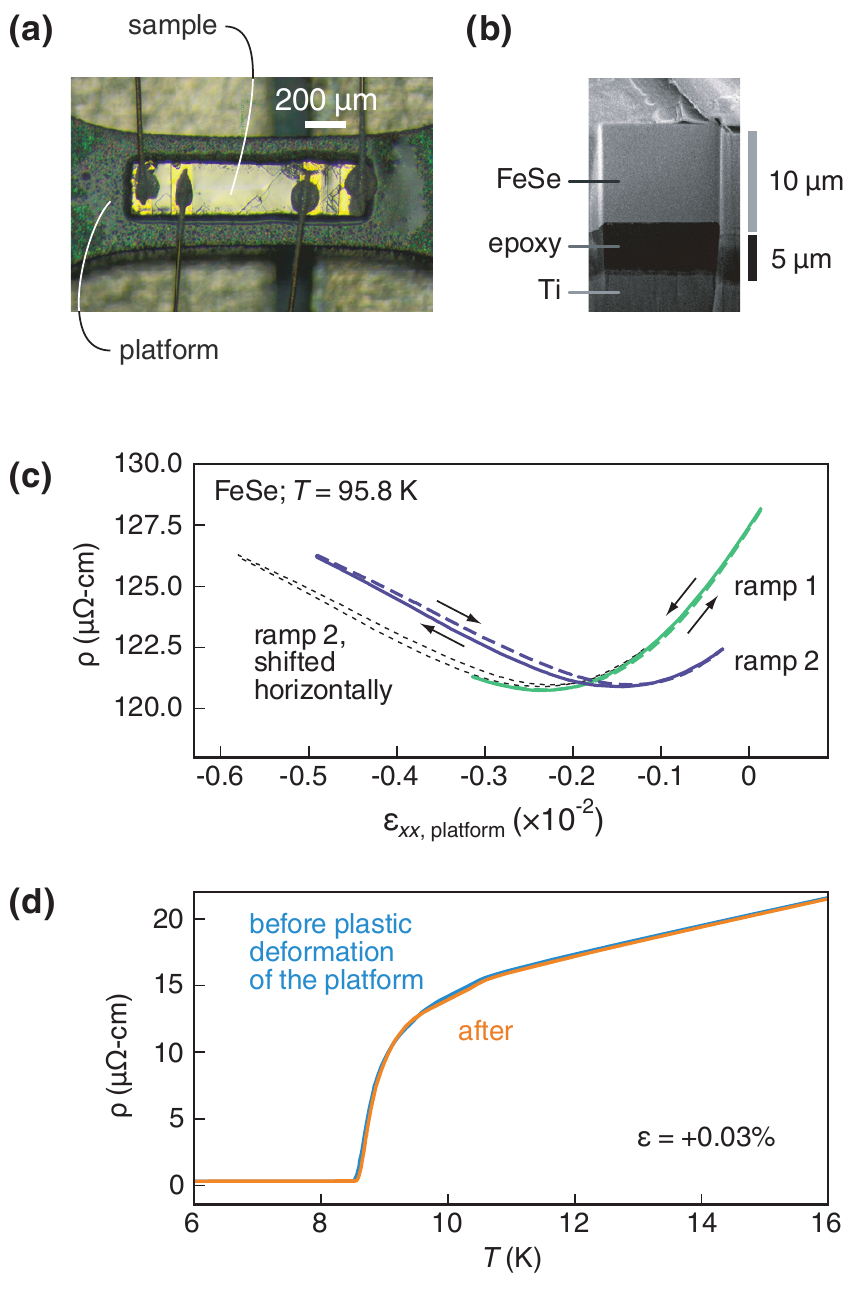}
\caption{\label{FeSeResults}\textbf{(a)} Photograph of a sample of FeSe mounted on an oxidized titanium platform. \textbf{(b)} A scanning electron microscopy image of a slice through sample, milled
with a focused ion beam, showing the sample and epoxy thickness. \textbf{(c)} Resistivity of this sample measured during strain ramps at 95.8~K. Driving the platform up to a strain of -0.58\% for the
second ramp caused the platform to deform plastically. This plastic deformation effectively locked in an additional transverse strain in the platform, shifting the observed $\rho(\varepsilon_{xx})$
curve horizontally. \textbf{(d)} The sample however did \textit{not} deform plastically: when the post-plastic-deformation strain was adjusted to match the $T_c$ observed beforehand, the
low-temperature resistance curve, shown here, was found to be essentially identical: no increase in resistivity due to introduction of dislocations, as was observed \textit{e.g.} when Sr$_2$RuO$_4$
was plastically deformed in Ref.~\onlinecite{Barber19}, was seen here.}
\end{figure}

A photograph of an FeSe sample mounted on a platform is shown in Fig.~\ref{FeSeResults}(a). Our mounting procedure was as follows. The single crystals were first cut into a bar shape using a wire saw.
Samples were then temporarily attached to a carrier plate using CrystalBond, and repeatedly cleaved using adhesive tape. In this way, thicknesses of less than 20~$\mu$m were achieved. To create stable
and low-resistance contacts, the surface was cleaned with a 10 minute plasma etch, and 150~nm of gold (without any adhesion layer) was sputtered onto the four contact regions.  The center of the
platform was then covered with a thin layer of MasterBond EP29LPSP epoxy, a low-viscosity epoxy, spread to a similar footprint as that of the sample. The sample was placed using static electricity
with a polymer-tipped tool made by MiTeGen. It was then gently pressed down using the same tool, before curing the epoxy at 70$^\circ$C for ten hours. This heating initially reduces the viscosity of
the epoxy, which wicks around the sample and forms smooth ramps along its edges.  Lastly, 25~$\mu$m diameter gold wires were attached using silver epoxy cured at room temperature.

This recipe gave epoxy layers of thickness 5--10~$\mu$m [Fig.~\ref{FeSeResults}(b)]. The samples were not flat on this scale, so this may be a lower limit set by the sample shape rather than the
viscosity of the epoxy. The sample photographed in Fig.~\ref{FeSeResults}(a) is 10~$\mu$m thick, so taking $C = C_{11} - C_{13}^2/C_{33} \sim 40$~GPa for FeSe gives a strain transmission length of
$\lambda \sim 40$~$\mu$m. This sample is 230~$\mu$m wide, and can therefore be taken to good approximation to be in the wide-sample limit. Fig.~\ref{FeSeResults}(c) shows resistivity versus applied
longitudinal strain of this sample, at temperature $T = 95.8$~K. The resistivity of FeSe has been shown to be very sensitive to anisotropic strain at low strains~\cite{Watson15, Hosoi16}; here we show
measurements up to much higher strains. In the first ramp, the strain was ramped from +0.02\% to -0.31\% and back, where negative values denote compression. The strain is taken as the applied
displacement divided by $l_\text{eff}$. In the second ramp, the strain was ramped from -0.58\% to -0.12\%, then back.

There is small hysteresis within each pair of curves [Fig.~\ref{FeSeResults}(c), blue and green], but very substantial offset between the two pairs [Fig.~\ref{FeSeResults}(c)]. This is a consequence
of plastic deformation of the platform: the elastic limit of the titanium of the platform at 95.8~K was exceeded when the strain was ramped to -0.58\%. This caused material in the platform neck to
``flow'' outward, carrying the sample with it and introducing, in effect, an offset in the anisotropic strain $\varepsilon_{xx} - \varepsilon_{yy}$. Upon reversing the direction of the strain ramp the
platform deformation was again elastic for some range, and the dominant effect of the offset introduced into $\varepsilon_{xx} - \varepsilon_{yy}$ was a horizontal offset between the low- and
high-strain strain ramps shown in Fig.~\ref{FeSeResults}(c).  Crucially, the sample did \textit{not} deform plastically: as shown in Fig.~\ref{FeSeResults}(d), to within resolution, the
low-temperature resistivity of the sample did not increase with the application of large strain, indicating that dislocations were not introduced into the sample. In other words, this method of sample
mounting can be used to apply elastic strains of at least 0.5\% to a mechanically delicate, van der Waals-bonded material such as FeSe. 

\begin{figure}[htbp]
\includegraphics[width=85mm]{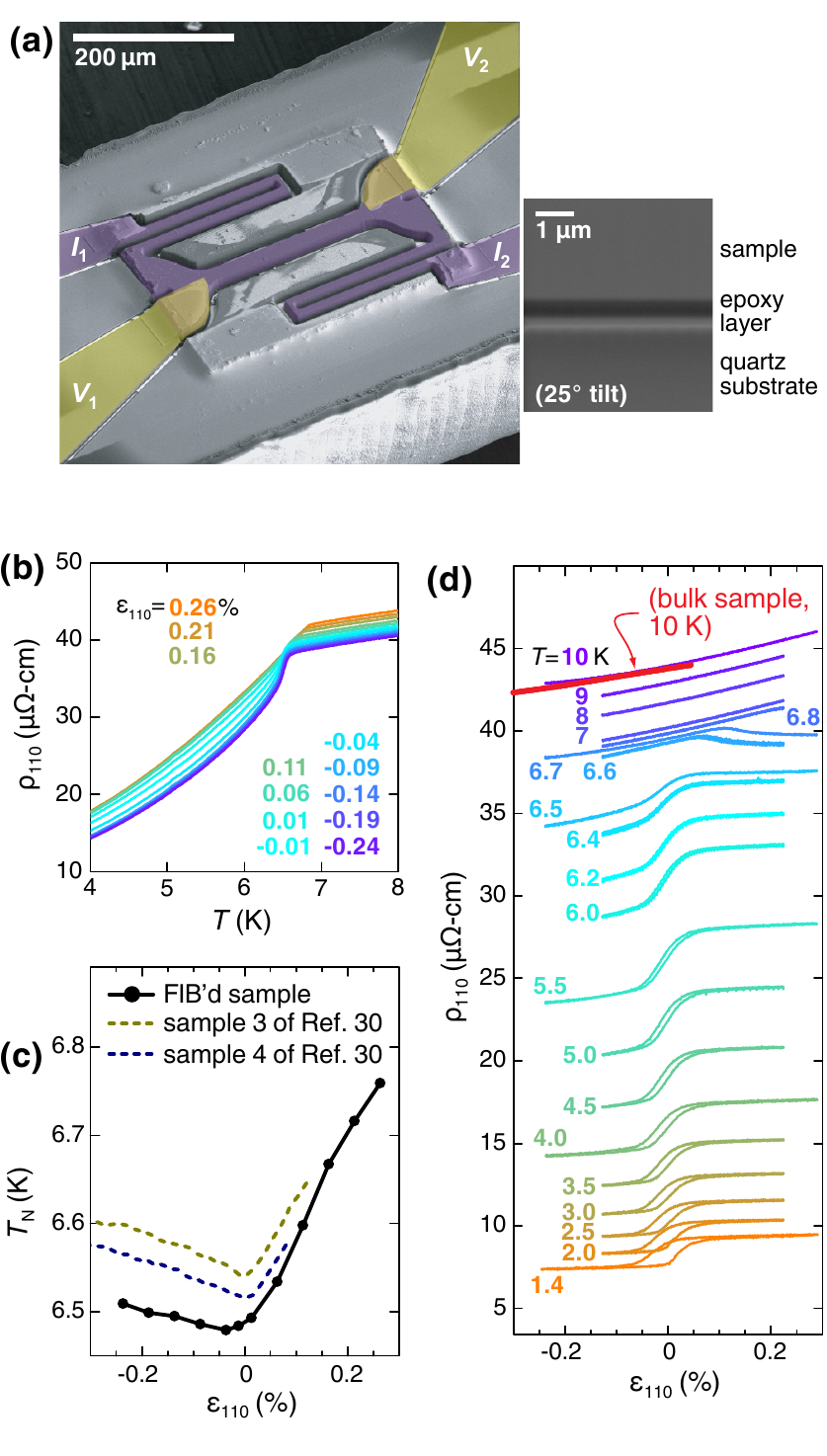}
\caption{\label{CeAuSb2Results}Results on CeAuSb$_2$. \textbf{(a)} Scanning electron micrograph of a sample of the heavy fermion antiferromagnet CeAuSb$_2$, affixed to a quartz platform and shaped
using a focused ion beam. The sample was oriented so that its long axis is along a $\langle 110 \rangle$ lattice direction. The current leads are colored purple and the voltage leads yellow. The panel
at right shows a micrograph of a slice through the neck region of the sample and into the quartz (done with the focused ion beam), showing that the epoxy layer between the sample and quartz was
$\sim$1~$\mu$m thick. \textbf{(b)} Resistivity $\rho$ versus temperature of this sample at various applied strains. \textbf{(c)} $T_N$, extracted from the data in panel (b), versus strain
$\varepsilon_{110}$. For comparison, results from measurement on two bulk samples, reported in Ref.~\onlinecite{Park18}, are shown. \textbf{(d)} $\rho$ versus $\varepsilon_{110}$ for the microstructured
sample at various temperatures. The heavy red line is the result from measurement on a bulk sample, reported in Ref.~\onlinecite{Park18}.}
\end{figure}

Fig.~\ref{CeAuSb2Results}(a) shows a sample of the heavy fermion antiferromagnet CeAuSb$_2$ mounted on a quartz platform and shaped with an ion beam. This particular sample incorporates long current
leads: with microstructured samples, the most practical way to deposit contacts is deposition from above, and the long leads allow the current to spread through the full thickness of the sample.

The CeAuSb$_2$ sample was prepared using the following procedure. First, the sample was polished to a thickness of $\sim$20~$\mu$m and then cut with a wire saw to dimensions of $300 \times
200$~$\mu$m. A 50 nm/ 300 nm composite layer of Ti/Au was deposited over the entire upper surface, and the sample was then mounted onto a quartz platform with Stycast 1266. The epoxy was mixed at room
temperature and degassed for 10 minutes in vacuum, and was then applied to the platform. Next, the epoxy was preheated on a hot plate to 65$^\circ$~C before placing the sample, in order to reduce its
viscosity. The epoxy was allowed to spread by capillary action after the sample was placed on top, without applying any additional force to the surface of the sample.  With some practice, we learned
to judge the size of the epoxy droplet so that in the end the epoxy thickness was $\sim$1~$\mu$m. The epoxy formed natural ramps up the edges of the sample. The epoxy was left to cure at 65$^\circ$~C
for about six hours, and then another layer of gold  was deposited to make connection to the sample via the epoxy ramps. The sample was then milled into the desired shape using a focused ion beam.

The elastic moduli of CeAuSb$_2$ have not been measured. Taking $E = 100$~GPa (a typical value for metals), $t = 20$~$\mu$m, $d = 1$~$\mu$m, and $G = 1.7$~GPa yields a strain transmission length of
$\lambda = 34$~$\mu$m, so the total length of this sample is $\sim 9 \lambda$. End tabs were incorporated into the sample shape, as described above, to aid strain transfer. The width of the neck, at
15~$\mu$m, is $\sim 0.4 \lambda$, so $\varepsilon_{yy}$ in the neck will be decoupled from that in the platform.

Results of measurement are shown in Fig.~\ref{CeAuSb2Results}(b-d). Again, strain is taken as applied displacement divided by $l_\text{eff}$. CeAuSb$_2$ has a transition into spin density wave
order at N{\'e}el temperature $T_N$=6.5(1)~K~\cite{Zhao16}, which can be clearly identified by a sharp drop in resistivity, as seen in panel (b). The propagation vectors of the spin density waves are
(0.136(2), $\pm$0.136(2), 0.5), where the $\pm$ indicates different domains~\cite{Marcus18}. As a result of domain formation, there is a first-order transition due to domain flipping as strain applied
along a $\langle 110 \rangle$ direction is ramped from compressive to tensile or vice versa, a process that has also been probed in bulk samples~\cite{Park18}. It manifests in two features in the
data. Firstly, when $T_N$ is plotted against strain applied along a $\langle 110 \rangle$ direction, $\varepsilon_{110}$, it shows an upward cusp at $\varepsilon_{110}=0$; this is shown in panel (c).
Secondly, when $\rho$ is measured against $\varepsilon_{110}$ below $T_N$, there is hysteresis across $\varepsilon_{110}=0$; this is shown in panel (d).

The results from the microstructured sample match well those recorded from bulk samples, demonstrating that rigid platforms can be used to pressurize microstructures. We note in addition that the
measurements on the microstructured sample extend to higher tensions than the bulk sample, as shown in panels (c) and (d).  The microstructured sample withstood higher tensions than the bulk samples,
likely because the ion beam milling leaves smoother edges than can be obtained from cutting with a wire saw, and so minimizing the appearance of edge defects at which stress concentrates, initiating
fractures.

\section{Conclusion}

We have described a method for applying anisotropic strain to samples by affixing them to platforms, and then applying uniaxial pressure to the platform using a piezoelectric-driven pressure cell. Key
to making this process work is to understand at the design stage the relative spring constants of the pressure cell and the platform. In the present case, the cell spring constant was 12~N/$\mu$m and
the platform spring constant 2--3~N/$\mu$m, ensuring that most of the displacement generated by the actuators went into deformation of the platform rather than the cell. Two platform materials were
demonstrated, fused quartz and titanium. 

This method allows large elastic strains to be applied to mechanically delicate samples. Here, an elastic strain of $\sim 5 \times 10^{-3}$ was demonstrated in the van der Waals-bonded material FeSe.
Attachment to a platform also facilitates shaping the sample with a focused ion beam, which was demonstrated here with a sample of CeAuSb$_2$. We anticipate a wide range of further platform-based
strain measurements.

\section*{Acknowledgment}
We acknowledge the financial support of the Max Planck Society. JP acknowledges the financial support of the National Research Foundation of Korea (NRF) funded by the Ministry of Science and ICT (No.
2016K1A4A4A01922028). Work in Japan was supported by a Grant-in-Aid for Scientific Research on Innovative Areas ``Quantum Liquid Crystals" (No. JP19H05824) from Japan Society for the Promotion of
Science. CWH has a 31\% ownership stake in Razorbill Instruments, a company that manufactures uniaxial pressure apparatus. Raw data for all figures in this paper are available at \textit{to be
determined}.

\section{Appendix}

\subsection{Discussion and measurement of the cell spring constant}

\begin{figure}[ptb]
\includegraphics[width=85mm]{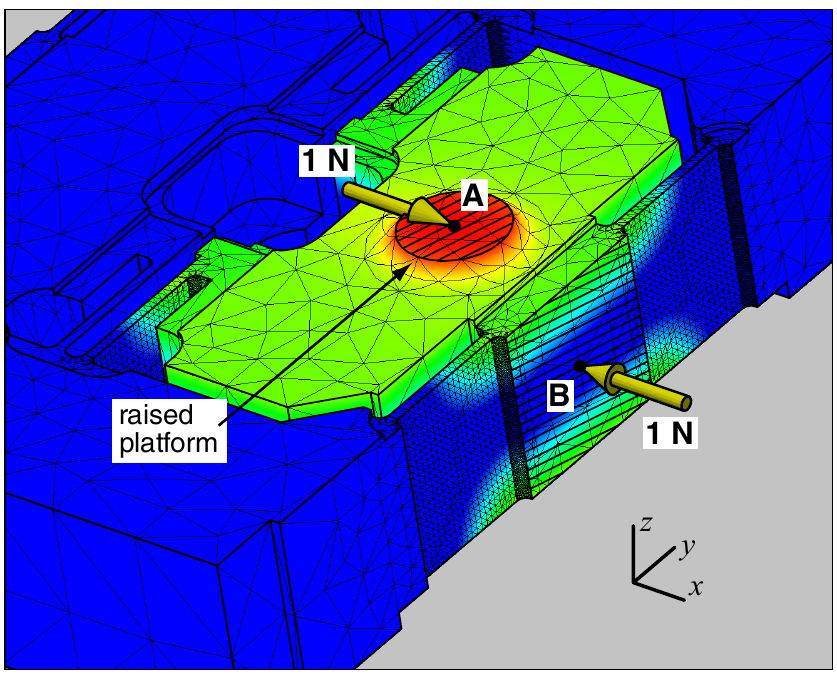}
\caption{\label{flexureDeformation}Finite element analysis of flexure deformation under the torque applied during operation of the cell. This analysis was done using Autodesk Inventor. Force is
applied across the hatched areas. The lower force represents the force applied by the actuators and the upper force the resisting force from the platform. The resisting force is applied to a raised
platform that is not present in the actual cell; this simulates the force being applied at a height of 0.5~mm above the upper surface of the cell. The model is colored by displacement along the $x$
axis; at point A it is 28.3~nm, and at point B, 0 nm. This gives a linear spring constant at point A of 1~N / 28.3~nm = 35~N/$\mu$m.}
\end{figure}

The metallic parts of the cell are made of titanium, which has a room-temperature Young's modulus of 103~GPa. In order to estimate the spring constant of the cell, we consider elastic deformation in
four areas. (1) The outer struts are slightly compressible; based on finite element analysis of these struts joined to the base plate, we estimate a spring constant for compression of the outer struts
of $\sim$230~N/$\mu$m. (2) The piezoelectric actuators have a room-temperature Young's modulus of $\approx$40~GPa, and the actuators each have dimensions $5 \times 5 \times 9$~mm$^3$. Mechanically,
the actuators labelled B in Fig.~1 are each in series with the two actuators on each side labelled A, which are in parallel with each other. The spring constant for compressing the actuators on one
side therefore comes to $\sim$74~N/$\mu$m. (3) The bridges that connect the actuators on each side bend slightly. The spring constant for bending a single bridge at the attachment points of the
actuators is $\sim$95~N/$\mu$m. (4) As described in the main text, during operation of the cell substantial torque is applied to the moving blocks, because the axis of the actuators is not aligned
with the axis of the platform. The flexures resist this torque, however not with infinite spring constant. Finite element analysis, illustrated in Fig.~\ref{flexureDeformation}, yields a spring
constant for rotation of a single moving block, as seen at a height 0.5~mm above the upper surface of the cell, of 35~N/$\mu$m. We note that this simulation neglects any contribution to rotational
stiffness from the piezoelectric actuators; including this contribution would increase the spring constant.  These separate spring constants can all be combined in series: $k_\text{cell}^{-1} = \sum
k_i^{-1}$, where $k_i$ is the spring constant of each element described above. This gives $k_\text{cell} = 9$~N/$\mu$m.

\begin{figure}[ptb]
\includegraphics[width=85mm]{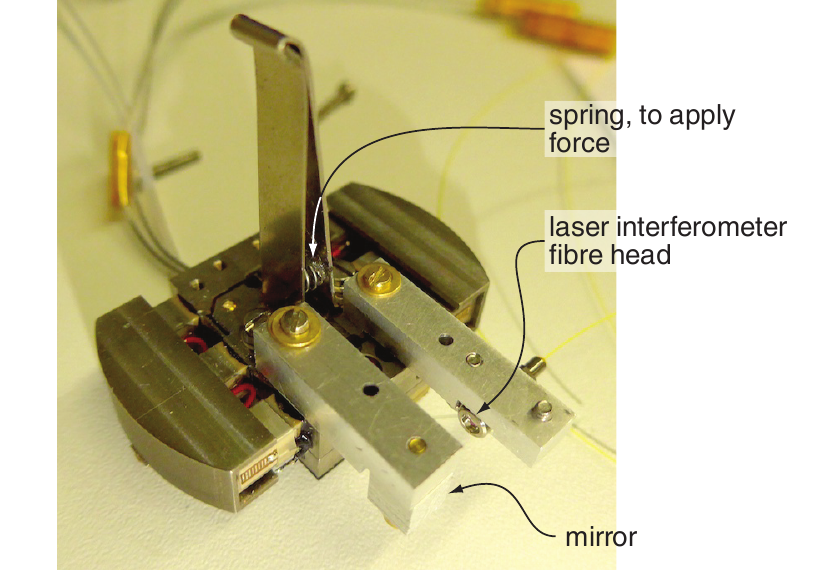}
\caption{\label{springConstantMeasurement}Configuration for measuring the spring constant of the cell, described further in the Appendix text.}
\end{figure}

Our setup for measuring the cell spring constant is shown in Fig.~\ref{springConstantMeasurement}. A fiber head of a laser interferometer were secured mechanically to the cell, and positioned so that
it was centered 0.5~mm above the surface of the cell. A spring was then inserted between two screws attached to the moving blocks, configured with aluminium levers so that the force would also be
applied at a height $\sim$0.5~mm above the surface of the cell. The force applied by the spring divided by the length change observed with the interferometer yielded the spring constant of the cell, as
seen for samples 0.5~mm above the upper surface of the cell: 12~N/$\mu$m. We note that because the platforms described here are mounted directly on the upper surface of the cell, they will see a
marginally higher cell spring constant.

\subsection{Torsional loading of the moving blocks}

Rotation of the moving blocks will also introduce a bending moment on the platform, however we show here that it is negligible. In the simulation shown in Fig.~\ref{flexureDeformation}, the torque
applied to the moving block is 3.2~N-mm, and the resulting rotation 28.3~nm/3.2~mm = $8.8 \times 10^{-6}$, indicating a torsional stiffness of the moving block of $k_\tau = 360$~N-m. This is an
underestimate, as the simulation neglects any contribution to torsional stiffness from the bending stiffness of the actuators.

Reaching, for example, a strain of $5 \times 10^{-3}$ in a titanium platform requires a force of 103~GPa $\times$ $5 \times 10^{-3}$ $\times$ 500~$\mu$m $\times$ 200~$\mu$m = 52~N. The platform is
centered 2.8~mm above the axis of the actuators, so the torque on each block is 146~N-mm, and the resulting rotation angle of each block $4.0 \times 10^{-4}$~rad. The total bend angle across the
platform is double this, because both blocks rotate, so the radius of curvature of the platform is $\approx l_\text{eff} / 8.1 \times 10^{-4} \approx 4.7$~m. This gives $\Delta \varepsilon / \langle
\varepsilon \rangle \approx 8 \times 10^{-3}$, where $\Delta \varepsilon$ is the difference in strain between the upper and lower surfaces of the platform, and $\langle \varepsilon \rangle = -5 \times
10^{-3}$ is the average strain in the center of the platform. The bending-induced strain gradient in the samples will be even smaller than this, because the samples are thinner.

The capacitive displacement sensor is centered 1.5~mm below the platform, so this rotation causes a 6\% difference between the displacement measured by this sensor and that actually applied to the
platform. However if $l_\text{eff}$ is calibrated using the displacement sensor in the cell then this discrepancy is included in the calibration. 

\subsection{Platform design considerations}

There are several variables to consider in designing the platform. 
\begin{enumerate}
\item The neck should be wide enough for practical sample mounting; we chose here a neck width of 0.5~mm. 
\item The cross-sectional area of the neck should substantially exceed that of samples that will be attached to it, so that the presence of the sample does not strongly affect the strain field within
the neck. 
\item As described in the main text, the combination of platform effective length, platform spring constant, and cell spring constant must be thought through at the design stage, to ensure that the
target strain can be reached. In the present case, we targeted a relatively short $l_\text{eff}$ and low platform spring constant, goals that in combination dictated a small cross section of the neck.
\item The platform must be thick enough not to buckle under the maximum strain desired in the measurement. 
\item The strain in the center of the neck should be relatively homogeneous. 
\item Stress concentration along the edges of the platform should be kept low, so that the achievable strain in the sample area is not limited by fracture or plastic deformation elsewhere in the
platform. In the present design, the maximum strain along the edge of the platform is 1.04 times the strain in the center of the neck.
\item When the platform and cell are made of different materials, $l_\text{eff}$ should be short enough that the actuators have enough range to compensate differential thermal contraction between the
cell and platform.
\end{enumerate}

\bibliography{literature_platforms}

\begin{thebibliography}{30}%
\makeatletter
\providecommand \@ifxundefined [1]{%
 \@ifx{#1\undefined}
}%
\providecommand \@ifnum [1]{%
 \ifnum #1\expandafter \@firstoftwo
 \else \expandafter \@secondoftwo
 \fi
}%
\providecommand \@ifx [1]{%
 \ifx #1\expandafter \@firstoftwo
 \else \expandafter \@secondoftwo
 \fi
}%
\providecommand \natexlab [1]{#1}%
\providecommand \enquote  [1]{``#1''}%
\providecommand \bibnamefont  [1]{#1}%
\providecommand \bibfnamefont [1]{#1}%
\providecommand \citenamefont [1]{#1}%
\providecommand \href@noop [0]{\@secondoftwo}%
\providecommand \href [0]{\begingroup \@sanitize@url \@href}%
\providecommand \@href[1]{\@@startlink{#1}\@@href}%
\providecommand \@@href[1]{\endgroup#1\@@endlink}%
\providecommand \@sanitize@url [0]{\catcode `\\12\catcode `\$12\catcode
  `\&12\catcode `\#12\catcode `\^12\catcode `\_12\catcode `\%12\relax}%
\providecommand \@@startlink[1]{}%
\providecommand \@@endlink[0]{}%
\providecommand \url  [0]{\begingroup\@sanitize@url \@url }%
\providecommand \@url [1]{\endgroup\@href {#1}{\urlprefix }}%
\providecommand \urlprefix  [0]{URL }%
\providecommand \Eprint [0]{\href }%
\providecommand \doibase [0]{http://dx.doi.org/}%
\providecommand \selectlanguage [0]{\@gobble}%
\providecommand \bibinfo  [0]{\@secondoftwo}%
\providecommand \bibfield  [0]{\@secondoftwo}%
\providecommand \translation [1]{[#1]}%
\providecommand \BibitemOpen [0]{}%
\providecommand \bibitemStop [0]{}%
\providecommand \bibitemNoStop [0]{.\EOS\space}%
\providecommand \EOS [0]{\spacefactor3000\relax}%
\providecommand \BibitemShut  [1]{\csname bibitem#1\endcsname}%
\let\auto@bib@innerbib\@empty
\bibitem [{\citenamefont {Steppke}\ \emph {et~al.}(2017)\citenamefont
  {Steppke}, \citenamefont {Zhao}, \citenamefont {Barber}, \citenamefont
  {Scaffidi}, \citenamefont {Jerzembeck}, \citenamefont {Rosner}, \citenamefont
  {Gibbs}, \citenamefont {Maeno}, \citenamefont {Simon}, \citenamefont
  {Mackenzie},\ and\ \citenamefont {Hicks}}]{Steppke17}%
  \BibitemOpen
  \bibfield  {author} {\bibinfo {author} {\bibfnamefont {A.}~\bibnamefont
  {Steppke}}, \bibinfo {author} {\bibfnamefont {L.}~\bibnamefont {Zhao}},
  \bibinfo {author} {\bibfnamefont {M.~E.}\ \bibnamefont {Barber}}, \bibinfo
  {author} {\bibfnamefont {T.}~\bibnamefont {Scaffidi}}, \bibinfo {author}
  {\bibfnamefont {F.}~\bibnamefont {Jerzembeck}}, \bibinfo {author}
  {\bibfnamefont {H.}~\bibnamefont {Rosner}}, \bibinfo {author} {\bibfnamefont
  {A.~S.}\ \bibnamefont {Gibbs}}, \bibinfo {author} {\bibfnamefont
  {Y.}~\bibnamefont {Maeno}}, \bibinfo {author} {\bibfnamefont {S.~H.}\
  \bibnamefont {Simon}}, \bibinfo {author} {\bibfnamefont {A.~P.}\ \bibnamefont
  {Mackenzie}}, \ and\ \bibinfo {author} {\bibfnamefont {C.~W.}\ \bibnamefont
  {Hicks}},\ }\href {\doibase 10.1126/science.aaf9398} {\bibfield  {journal}
  {\bibinfo  {journal} {Science}\ }\textbf {\bibinfo {volume} {355}},\ \bibinfo
  {pages} {eaaf9398} (\bibinfo {year} {2017})}\BibitemShut {NoStop}%
\bibitem [{\citenamefont {Shirakawa}\ \emph {et~al.}(1997)\citenamefont
  {Shirakawa}, \citenamefont {Murata}, \citenamefont {Nishizaki}, \citenamefont
  {Maeno},\ and\ \citenamefont {Fujita}}]{Shirakawa97}%
  \BibitemOpen
  \bibfield  {author} {\bibinfo {author} {\bibfnamefont {N.}~\bibnamefont
  {Shirakawa}}, \bibinfo {author} {\bibfnamefont {K.}~\bibnamefont {Murata}},
  \bibinfo {author} {\bibfnamefont {S.}~\bibnamefont {Nishizaki}}, \bibinfo
  {author} {\bibfnamefont {Y.}~\bibnamefont {Maeno}}, \ and\ \bibinfo {author}
  {\bibfnamefont {T.}~\bibnamefont {Fujita}},\ }\href {\doibase
  10.1103/PhysRevB.56.7890} {\bibfield  {journal} {\bibinfo  {journal}
  {Physical Review B}\ }\textbf {\bibinfo {volume} {56}},\ \bibinfo {pages}
  {7890} (\bibinfo {year} {1997})}\BibitemShut {NoStop}%
\bibitem [{\citenamefont {Souliou}\ \emph {et~al.}(2018)\citenamefont
  {Souliou}, \citenamefont {Gretarsson}, \citenamefont {Garbarino},
  \citenamefont {Bosak}, \citenamefont {Porras}, \citenamefont {Loew},
  \citenamefont {Keimer},\ and\ \citenamefont {Le~Tacon}}]{Souliou18}%
  \BibitemOpen
  \bibfield  {author} {\bibinfo {author} {\bibfnamefont {S.~M.}\ \bibnamefont
  {Souliou}}, \bibinfo {author} {\bibfnamefont {H.}~\bibnamefont {Gretarsson}},
  \bibinfo {author} {\bibfnamefont {G.}~\bibnamefont {Garbarino}}, \bibinfo
  {author} {\bibfnamefont {A.}~\bibnamefont {Bosak}}, \bibinfo {author}
  {\bibfnamefont {J.}~\bibnamefont {Porras}}, \bibinfo {author} {\bibfnamefont
  {T.}~\bibnamefont {Loew}}, \bibinfo {author} {\bibfnamefont {B.}~\bibnamefont
  {Keimer}}, \ and\ \bibinfo {author} {\bibfnamefont {M.}~\bibnamefont
  {Le~Tacon}},\ }\href {\doibase 10.1103/PhysRevB.97.020503} {\bibfield
  {journal} {\bibinfo  {journal} {Physical Review B}\ }\textbf {\bibinfo
  {volume} {97}},\ \bibinfo {pages} {020503} (\bibinfo {year}
  {2018})}\BibitemShut {NoStop}%
\bibitem [{\citenamefont {Kim}\ \emph {et~al.}(2018)\citenamefont {Kim},
  \citenamefont {Souliou}, \citenamefont {Barber}, \citenamefont {Lefrançois},
  \citenamefont {Minola}, \citenamefont {Tortora}, \citenamefont {Heid},
  \citenamefont {Nandi}, \citenamefont {Borzi}, \citenamefont {Garbarino},
  \citenamefont {Bosak}, \citenamefont {Porras}, \citenamefont {Loew},
  \citenamefont {König}, \citenamefont {Moll}, \citenamefont {Mackenzie},
  \citenamefont {Keimer}, \citenamefont {Hicks},\ and\ \citenamefont
  {Tacon}}]{Kim18}%
  \BibitemOpen
  \bibfield  {author} {\bibinfo {author} {\bibfnamefont {H.-H.}\ \bibnamefont
  {Kim}}, \bibinfo {author} {\bibfnamefont {S.~M.}\ \bibnamefont {Souliou}},
  \bibinfo {author} {\bibfnamefont {M.~E.}\ \bibnamefont {Barber}}, \bibinfo
  {author} {\bibfnamefont {E.}~\bibnamefont {Lefrançois}}, \bibinfo {author}
  {\bibfnamefont {M.}~\bibnamefont {Minola}}, \bibinfo {author} {\bibfnamefont
  {M.}~\bibnamefont {Tortora}}, \bibinfo {author} {\bibfnamefont
  {R.}~\bibnamefont {Heid}}, \bibinfo {author} {\bibfnamefont {N.}~\bibnamefont
  {Nandi}}, \bibinfo {author} {\bibfnamefont {R.~A.}\ \bibnamefont {Borzi}},
  \bibinfo {author} {\bibfnamefont {G.}~\bibnamefont {Garbarino}}, \bibinfo
  {author} {\bibfnamefont {A.}~\bibnamefont {Bosak}}, \bibinfo {author}
  {\bibfnamefont {J.}~\bibnamefont {Porras}}, \bibinfo {author} {\bibfnamefont
  {T.}~\bibnamefont {Loew}}, \bibinfo {author} {\bibfnamefont {M.}~\bibnamefont
  {König}}, \bibinfo {author} {\bibfnamefont {P.~J.~W.}\ \bibnamefont {Moll}},
  \bibinfo {author} {\bibfnamefont {A.~P.}\ \bibnamefont {Mackenzie}}, \bibinfo
  {author} {\bibfnamefont {B.}~\bibnamefont {Keimer}}, \bibinfo {author}
  {\bibfnamefont {C.~W.}\ \bibnamefont {Hicks}}, \ and\ \bibinfo {author}
  {\bibfnamefont {M.~L.}\ \bibnamefont {Tacon}},\ }\href {\doibase
  10.1126/science.aat4708} {\bibfield  {journal} {\bibinfo  {journal}
  {Science}\ }\textbf {\bibinfo {volume} {362}},\ \bibinfo {pages} {1040}
  (\bibinfo {year} {2018})}\BibitemShut {NoStop}%
\bibitem [{\citenamefont {Kuo}\ \emph {et~al.}(2016)\citenamefont {Kuo},
  \citenamefont {Chu}, \citenamefont {Palmstrom}, \citenamefont {Kivelson},\
  and\ \citenamefont {Fisher}}]{Kuo16}%
  \BibitemOpen
  \bibfield  {author} {\bibinfo {author} {\bibfnamefont {H.-H.}\ \bibnamefont
  {Kuo}}, \bibinfo {author} {\bibfnamefont {J.-H.}\ \bibnamefont {Chu}},
  \bibinfo {author} {\bibfnamefont {J.~C.}\ \bibnamefont {Palmstrom}}, \bibinfo
  {author} {\bibfnamefont {S.~A.}\ \bibnamefont {Kivelson}}, \ and\ \bibinfo
  {author} {\bibfnamefont {I.~R.}\ \bibnamefont {Fisher}},\ }\href {\doibase
  10.1126/science.aab0103} {\bibfield  {journal} {\bibinfo  {journal}
  {Science}\ }\textbf {\bibinfo {volume} {352}},\ \bibinfo {pages} {958}
  (\bibinfo {year} {2016})}\BibitemShut {NoStop}%
\bibitem [{\citenamefont {Hicks}\ \emph
  {et~al.}(2014{\natexlab{a}})\citenamefont {Hicks}, \citenamefont {Brodsky},
  \citenamefont {Yelland}, \citenamefont {Gibbs}, \citenamefont {Bruin},
  \citenamefont {Barber}, \citenamefont {Edkins}, \citenamefont {Nishimura},
  \citenamefont {Yonezawa}, \citenamefont {Maeno} \emph {et~al.}}]{Hicks14}%
  \BibitemOpen
  \bibfield  {author} {\bibinfo {author} {\bibfnamefont {C.~W.}\ \bibnamefont
  {Hicks}}, \bibinfo {author} {\bibfnamefont {D.~O.}\ \bibnamefont {Brodsky}},
  \bibinfo {author} {\bibfnamefont {E.~A.}\ \bibnamefont {Yelland}}, \bibinfo
  {author} {\bibfnamefont {A.~S.}\ \bibnamefont {Gibbs}}, \bibinfo {author}
  {\bibfnamefont {J.~A.}\ \bibnamefont {Bruin}}, \bibinfo {author}
  {\bibfnamefont {M.~E.}\ \bibnamefont {Barber}}, \bibinfo {author}
  {\bibfnamefont {S.~D.}\ \bibnamefont {Edkins}}, \bibinfo {author}
  {\bibfnamefont {K.}~\bibnamefont {Nishimura}}, \bibinfo {author}
  {\bibfnamefont {S.}~\bibnamefont {Yonezawa}}, \bibinfo {author}
  {\bibfnamefont {Y.}~\bibnamefont {Maeno}},  \emph {et~al.},\ }\href@noop {}
  {\bibfield  {journal} {\bibinfo  {journal} {Science}\ }\textbf {\bibinfo
  {volume} {344}},\ \bibinfo {pages} {283} (\bibinfo {year}
  {2014}{\natexlab{a}})}\BibitemShut {NoStop}%
\bibitem [{\citenamefont {Hicks}\ \emph
  {et~al.}(2014{\natexlab{b}})\citenamefont {Hicks}, \citenamefont {Barber},
  \citenamefont {Edkins}, \citenamefont {Brodsky},\ and\ \citenamefont
  {Mackenzie}}]{Hicks14RSI}%
  \BibitemOpen
  \bibfield  {author} {\bibinfo {author} {\bibfnamefont {C.~W.}\ \bibnamefont
  {Hicks}}, \bibinfo {author} {\bibfnamefont {M.~E.}\ \bibnamefont {Barber}},
  \bibinfo {author} {\bibfnamefont {S.~D.}\ \bibnamefont {Edkins}}, \bibinfo
  {author} {\bibfnamefont {D.~O.}\ \bibnamefont {Brodsky}}, \ and\ \bibinfo
  {author} {\bibfnamefont {A.~P.}\ \bibnamefont {Mackenzie}},\ }\href {\doibase
  10.1063/1.4881611} {\bibfield  {journal} {\bibinfo  {journal} {Review of
  Scientific Instruments}\ }\textbf {\bibinfo {volume} {85}},\ \bibinfo {pages}
  {065003} (\bibinfo {year} {2014}{\natexlab{b}})}\BibitemShut {NoStop}%
\bibitem [{\citenamefont {Barber}\ \emph {et~al.}(2019)\citenamefont {Barber},
  \citenamefont {Steppke}, \citenamefont {Mackenzie},\ and\ \citenamefont
  {Hicks}}]{Barber19}%
  \BibitemOpen
  \bibfield  {author} {\bibinfo {author} {\bibfnamefont {M.~E.}\ \bibnamefont
  {Barber}}, \bibinfo {author} {\bibfnamefont {A.}~\bibnamefont {Steppke}},
  \bibinfo {author} {\bibfnamefont {A.~P.}\ \bibnamefont {Mackenzie}}, \ and\
  \bibinfo {author} {\bibfnamefont {C.~W.}\ \bibnamefont {Hicks}},\ }\href
  {\doibase 10.1063/1.5075485} {\bibfield  {journal} {\bibinfo  {journal}
  {Review of Scientific Instruments}\ }\textbf {\bibinfo {volume} {90}},\
  \bibinfo {pages} {023904} (\bibinfo {year} {2019})}\BibitemShut {NoStop}%
\bibitem [{\citenamefont {Kissikov}\ \emph {et~al.}(2018)\citenamefont
  {Kissikov}, \citenamefont {Sarkar}, \citenamefont {Lawson}, \citenamefont
  {Bush}, \citenamefont {Timmons}, \citenamefont {Tanatar}, \citenamefont
  {Prozorov}, \citenamefont {Bud’ko}, \citenamefont {Canfield}, \citenamefont
  {Fernandes} \emph {et~al.}}]{Kissikov18}%
  \BibitemOpen
  \bibfield  {author} {\bibinfo {author} {\bibfnamefont {T.}~\bibnamefont
  {Kissikov}}, \bibinfo {author} {\bibfnamefont {R.}~\bibnamefont {Sarkar}},
  \bibinfo {author} {\bibfnamefont {M.}~\bibnamefont {Lawson}}, \bibinfo
  {author} {\bibfnamefont {B.}~\bibnamefont {Bush}}, \bibinfo {author}
  {\bibfnamefont {E.~I.}\ \bibnamefont {Timmons}}, \bibinfo {author}
  {\bibfnamefont {M.~A.}\ \bibnamefont {Tanatar}}, \bibinfo {author}
  {\bibfnamefont {R.}~\bibnamefont {Prozorov}}, \bibinfo {author}
  {\bibfnamefont {S.}~\bibnamefont {Bud’ko}}, \bibinfo {author}
  {\bibfnamefont {P.~C.}\ \bibnamefont {Canfield}}, \bibinfo {author}
  {\bibfnamefont {R.}~\bibnamefont {Fernandes}},  \emph {et~al.},\ }\href@noop
  {} {\bibfield  {journal} {\bibinfo  {journal} {Nature communications}\
  }\textbf {\bibinfo {volume} {9}},\ \bibinfo {pages} {1058} (\bibinfo {year}
  {2018})}\BibitemShut {NoStop}%
\bibitem [{\citenamefont {Ikeda}\ \emph {et~al.}(2018)\citenamefont {Ikeda},
  \citenamefont {Worasaran}, \citenamefont {Palmstrom}, \citenamefont
  {Straquadine}, \citenamefont {Walmsley},\ and\ \citenamefont
  {Fisher}}]{Ikeda18}%
  \BibitemOpen
  \bibfield  {author} {\bibinfo {author} {\bibfnamefont {M.~S.}\ \bibnamefont
  {Ikeda}}, \bibinfo {author} {\bibfnamefont {T.}~\bibnamefont {Worasaran}},
  \bibinfo {author} {\bibfnamefont {J.~C.}\ \bibnamefont {Palmstrom}}, \bibinfo
  {author} {\bibfnamefont {J.}~\bibnamefont {Straquadine}}, \bibinfo {author}
  {\bibfnamefont {P.}~\bibnamefont {Walmsley}}, \ and\ \bibinfo {author}
  {\bibfnamefont {I.}~\bibnamefont {Fisher}},\ }\href@noop {} {\bibfield
  {journal} {\bibinfo  {journal} {Physical Review B}\ }\textbf {\bibinfo
  {volume} {98}},\ \bibinfo {pages} {245133} (\bibinfo {year}
  {2018})}\BibitemShut {NoStop}%
\bibitem [{\citenamefont {Mutch}\ \emph {et~al.}(2019)\citenamefont {Mutch},
  \citenamefont {Chen}, \citenamefont {Went}, \citenamefont {Qian},
  \citenamefont {Wilson}, \citenamefont {Andreev}, \citenamefont {Chen},\ and\
  \citenamefont {Chu}}]{Mutch19}%
  \BibitemOpen
  \bibfield  {author} {\bibinfo {author} {\bibfnamefont {J.}~\bibnamefont
  {Mutch}}, \bibinfo {author} {\bibfnamefont {W.-C.}\ \bibnamefont {Chen}},
  \bibinfo {author} {\bibfnamefont {P.}~\bibnamefont {Went}}, \bibinfo {author}
  {\bibfnamefont {T.}~\bibnamefont {Qian}}, \bibinfo {author} {\bibfnamefont
  {I.~Z.}\ \bibnamefont {Wilson}}, \bibinfo {author} {\bibfnamefont
  {A.}~\bibnamefont {Andreev}}, \bibinfo {author} {\bibfnamefont {C.-C.}\
  \bibnamefont {Chen}}, \ and\ \bibinfo {author} {\bibfnamefont {J.-H.}\
  \bibnamefont {Chu}},\ }\href@noop {} {\bibfield  {journal} {\bibinfo
  {journal} {Science advances}\ }\textbf {\bibinfo {volume} {5}},\ \bibinfo
  {pages} {eaav9771} (\bibinfo {year} {2019})}\BibitemShut {NoStop}%
\bibitem [{\citenamefont {Chu}\ \emph {et~al.}(2012)\citenamefont {Chu},
  \citenamefont {Kuo}, \citenamefont {Analytis},\ and\ \citenamefont
  {Fisher}}]{Chu12}%
  \BibitemOpen
  \bibfield  {author} {\bibinfo {author} {\bibfnamefont {J.-H.}\ \bibnamefont
  {Chu}}, \bibinfo {author} {\bibfnamefont {H.-H.}\ \bibnamefont {Kuo}},
  \bibinfo {author} {\bibfnamefont {J.~G.}\ \bibnamefont {Analytis}}, \ and\
  \bibinfo {author} {\bibfnamefont {I.~R.}\ \bibnamefont {Fisher}},\ }\href
  {\doibase 10.1126/science.1221713} {\bibfield  {journal} {\bibinfo  {journal}
  {Science}\ }\textbf {\bibinfo {volume} {337}},\ \bibinfo {pages} {710}
  (\bibinfo {year} {2012})}\BibitemShut {NoStop}%
\bibitem [{\citenamefont {Moll}\ \emph {et~al.}(2016)\citenamefont {Moll},
  \citenamefont {Kushwaha}, \citenamefont {Nandi}, \citenamefont {Schmidt},\
  and\ \citenamefont {Mackenzie}}]{Moll16}%
  \BibitemOpen
  \bibfield  {author} {\bibinfo {author} {\bibfnamefont {P.~J.}\ \bibnamefont
  {Moll}}, \bibinfo {author} {\bibfnamefont {P.}~\bibnamefont {Kushwaha}},
  \bibinfo {author} {\bibfnamefont {N.}~\bibnamefont {Nandi}}, \bibinfo
  {author} {\bibfnamefont {B.}~\bibnamefont {Schmidt}}, \ and\ \bibinfo
  {author} {\bibfnamefont {A.~P.}\ \bibnamefont {Mackenzie}},\ }\href@noop {}
  {\bibfield  {journal} {\bibinfo  {journal} {Science}\ }\textbf {\bibinfo
  {volume} {351}},\ \bibinfo {pages} {1061} (\bibinfo {year}
  {2016})}\BibitemShut {NoStop}%
\bibitem [{\citenamefont {Rosenberg}\ \emph {et~al.}(2019)\citenamefont
  {Rosenberg}, \citenamefont {Chu}, \citenamefont {Ruff}, \citenamefont
  {Hristov},\ and\ \citenamefont {Fisher}}]{Rosenberg19}%
  \BibitemOpen
  \bibfield  {author} {\bibinfo {author} {\bibfnamefont {E.~W.}\ \bibnamefont
  {Rosenberg}}, \bibinfo {author} {\bibfnamefont {J.-H.}\ \bibnamefont {Chu}},
  \bibinfo {author} {\bibfnamefont {J.~P.~C.}\ \bibnamefont {Ruff}}, \bibinfo
  {author} {\bibfnamefont {A.~T.}\ \bibnamefont {Hristov}}, \ and\ \bibinfo
  {author} {\bibfnamefont {I.~R.}\ \bibnamefont {Fisher}},\ }\href {\doibase
  10.1073/pnas.1818910116} {\bibfield  {journal} {\bibinfo  {journal}
  {Proceedings of the National Academy of Sciences}\ }\textbf {\bibinfo
  {volume} {116}},\ \bibinfo {pages} {7232} (\bibinfo {year} {2019})},\ \Eprint
  {http://arxiv.org/abs/https://www.pnas.org/content/116/15/7232.full.pdf}
  {https://www.pnas.org/content/116/15/7232.full.pdf} \BibitemShut {NoStop}%
\bibitem [{\citenamefont {Kostylev}, \citenamefont {Yonezawa},\ and\
  \citenamefont {Maeno}(2019)}]{Kostylev19}%
  \BibitemOpen
  \bibfield  {author} {\bibinfo {author} {\bibfnamefont {I.}~\bibnamefont
  {Kostylev}}, \bibinfo {author} {\bibfnamefont {S.}~\bibnamefont {Yonezawa}},
  \ and\ \bibinfo {author} {\bibfnamefont {Y.}~\bibnamefont {Maeno}},\
  }\href@noop {} {\bibfield  {journal} {\bibinfo  {journal} {Journal of Applied
  Physics}\ }\textbf {\bibinfo {volume} {125}},\ \bibinfo {pages} {082535}
  (\bibinfo {year} {2019})}\BibitemShut {NoStop}%
\bibitem [{Note1()}]{Note1}%
  \BibitemOpen
  \bibinfo {note} {Goodfellow, grade 2 (chemistry only) temper annealed
  titanium foils (Part No. TI000400), post-processed by KMLT Dermicut GmbH,
  Neukirch, Germany.}\BibitemShut {Stop}%
\bibitem [{Note2()}]{Note2}%
  \BibitemOpen
  \bibinfo {note} {Plan Optik AG, Eisoff, Germany, 4 inch-diameter, 200~$\mu
  $m-thick quartz wafers, post-processed by Lasercut24, Golmsdorf,
  Germany.}\BibitemShut {Stop}%
\bibitem [{Note3()}]{Note3}%
  \BibitemOpen
  \bibinfo {note} {The actuators used in this cell are PICMA\protect
  \textsuperscript {\textregistered } actuators from Physik Instrumente, model
  number P885.10. Their response at 1.5~K is reported in Ref.~\cite
  {Barber19}.}\BibitemShut {Stop}%
\bibitem [{\citenamefont {Okaji}\ \emph {et~al.}(1995)\citenamefont {Okaji},
  \citenamefont {Yamada}, \citenamefont {Nara},\ and\ \citenamefont
  {Kato}}]{Okaji95}%
  \BibitemOpen
  \bibfield  {author} {\bibinfo {author} {\bibfnamefont {M.}~\bibnamefont
  {Okaji}}, \bibinfo {author} {\bibfnamefont {N.}~\bibnamefont {Yamada}},
  \bibinfo {author} {\bibfnamefont {K.}~\bibnamefont {Nara}}, \ and\ \bibinfo
  {author} {\bibfnamefont {H.}~\bibnamefont {Kato}},\ }\href@noop {} {\bibfield
   {journal} {\bibinfo  {journal} {Cryogenics}\ }\textbf {\bibinfo {volume}
  {35}},\ \bibinfo {pages} {887} (\bibinfo {year} {1995})}\BibitemShut
  {NoStop}%
\bibitem [{\citenamefont {Turner}, \citenamefont {Crozier},\ and\ \citenamefont
  {Reu}(2015)}]{Turner2015}%
  \BibitemOpen
  \bibfield  {author} {\bibinfo {author} {\bibfnamefont {D.}~\bibnamefont
  {Turner}}, \bibinfo {author} {\bibfnamefont {P.}~\bibnamefont {Crozier}}, \
  and\ \bibinfo {author} {\bibfnamefont {P.}~\bibnamefont {Reu}},\ }\href
  {https://www.osti.gov//servlets/purl/1245432} {\enquote {\bibinfo {title}
  {{Digital Image Correlation Engine (DICe), Version 2.0}},}\ } (\bibinfo
  {year} {2015})\BibitemShut {NoStop}%
\bibitem [{\citenamefont {Sunko}\ \emph {et~al.}(2019)\citenamefont {Sunko},
  \citenamefont {Abarca~Morales}, \citenamefont {Markovi\'{c}}, \citenamefont
  {Barber}, \citenamefont {Milosavljevi\'{c}}, \citenamefont {Mazzola},
  \citenamefont {Sokolov}, \citenamefont {Kikugawa}, \citenamefont {Cacho},
  \citenamefont {Dudin}, \citenamefont {Rosner}, \citenamefont {Hicks},
  \citenamefont {King},\ and\ \citenamefont {Mackenzie}}]{Sunko19}%
  \BibitemOpen
  \bibfield  {author} {\bibinfo {author} {\bibfnamefont {V.}~\bibnamefont
  {Sunko}}, \bibinfo {author} {\bibfnamefont {E.}~\bibnamefont
  {Abarca~Morales}}, \bibinfo {author} {\bibfnamefont {I.}~\bibnamefont
  {Markovi\'{c}}}, \bibinfo {author} {\bibfnamefont {M.~E.}\ \bibnamefont
  {Barber}}, \bibinfo {author} {\bibfnamefont {D.}~\bibnamefont
  {Milosavljevi\'{c}}}, \bibinfo {author} {\bibfnamefont {F.}~\bibnamefont
  {Mazzola}}, \bibinfo {author} {\bibfnamefont {D.~A.}\ \bibnamefont
  {Sokolov}}, \bibinfo {author} {\bibfnamefont {N.}~\bibnamefont {Kikugawa}},
  \bibinfo {author} {\bibfnamefont {C.}~\bibnamefont {Cacho}}, \bibinfo
  {author} {\bibfnamefont {P.}~\bibnamefont {Dudin}}, \bibinfo {author}
  {\bibfnamefont {H.}~\bibnamefont {Rosner}}, \bibinfo {author} {\bibfnamefont
  {C.~W.}\ \bibnamefont {Hicks}}, \bibinfo {author} {\bibfnamefont {P.~D.~C.}\
  \bibnamefont {King}}, \ and\ \bibinfo {author} {\bibfnamefont {A.~P.}\
  \bibnamefont {Mackenzie}},\ }\href@noop {} {\bibfield  {journal} {\bibinfo
  {journal} {npj Quant. Materials}\ }\textbf {\bibinfo {volume} {4}},\ \bibinfo
  {pages} {46} (\bibinfo {year} {2019})}\BibitemShut {NoStop}%
\bibitem [{\citenamefont {He}\ \emph {et~al.}(2018)\citenamefont {He},
  \citenamefont {Wang}, \citenamefont {Ahn}, \citenamefont {Hardy},
  \citenamefont {Wolf}, \citenamefont {Adelmann}, \citenamefont {Schmalian},
  \citenamefont {Eremin},\ and\ \citenamefont {Maingast}}]{He18}%
  \BibitemOpen
  \bibfield  {author} {\bibinfo {author} {\bibfnamefont {M.~Q.}\ \bibnamefont
  {He}}, \bibinfo {author} {\bibfnamefont {L.~R.}\ \bibnamefont {Wang}},
  \bibinfo {author} {\bibfnamefont {F.}~\bibnamefont {Ahn}}, \bibinfo {author}
  {\bibfnamefont {F.}~\bibnamefont {Hardy}}, \bibinfo {author} {\bibfnamefont
  {T.}~\bibnamefont {Wolf}}, \bibinfo {author} {\bibfnamefont {P.}~\bibnamefont
  {Adelmann}}, \bibinfo {author} {\bibfnamefont {J.}~\bibnamefont {Schmalian}},
  \bibinfo {author} {\bibfnamefont {I.}~\bibnamefont {Eremin}}, \ and\ \bibinfo
  {author} {\bibfnamefont {C.}~\bibnamefont {Maingast}},\ }\href@noop {}
  {\bibfield  {journal} {\bibinfo  {journal} {Nature Communications}\ }\textbf
  {\bibinfo {volume} {8}},\ \bibinfo {pages} {504} (\bibinfo {year}
  {2018})}\BibitemShut {NoStop}%
\bibitem [{\citenamefont {Hashimoto}\ and\ \citenamefont
  {Ikushima}(1980)}]{Hashimoto80}%
  \BibitemOpen
  \bibfield  {author} {\bibinfo {author} {\bibfnamefont {T.}~\bibnamefont
  {Hashimoto}}\ and\ \bibinfo {author} {\bibfnamefont {A.}~\bibnamefont
  {Ikushima}},\ }\href {\doibase 10.1063/1.1136224} {\bibfield  {journal}
  {\bibinfo  {journal} {Review of Scientific Instruments}\ }\textbf {\bibinfo
  {volume} {51}},\ \bibinfo {pages} {378} (\bibinfo {year} {1980})}\BibitemShut
  {NoStop}%
\bibitem [{\citenamefont {Millican}\ \emph {et~al.}(2009)\citenamefont
  {Millican}, \citenamefont {Phelan}, \citenamefont {Thomas}, \citenamefont
  {Le\~{a}o},\ and\ \citenamefont {Carpenter}}]{Millican09}%
  \BibitemOpen
  \bibfield  {author} {\bibinfo {author} {\bibfnamefont {J.~N.}\ \bibnamefont
  {Millican}}, \bibinfo {author} {\bibfnamefont {D.}~\bibnamefont {Phelan}},
  \bibinfo {author} {\bibfnamefont {E.~L.}\ \bibnamefont {Thomas}}, \bibinfo
  {author} {\bibfnamefont {J.~B.}\ \bibnamefont {Le\~{a}o}}, \ and\ \bibinfo
  {author} {\bibfnamefont {E.}~\bibnamefont {Carpenter}},\ }\href@noop {}
  {\bibfield  {journal} {\bibinfo  {journal} {Solid State Communications}\
  }\textbf {\bibinfo {volume} {149}},\ \bibinfo {pages} {707} (\bibinfo {year}
  {2009})}\BibitemShut {NoStop}%
\bibitem [{\citenamefont {Zvyagina}\ \emph {et~al.}(2013)\citenamefont
  {Zvyagina}, \citenamefont {Gaydamak}, \citenamefont {Zhekov}, \citenamefont
  {Bilich}, \citenamefont {Fil}, \citenamefont {Chareev},\ and\ \citenamefont
  {Vasiliev}}]{Zvyagina13}%
  \BibitemOpen
  \bibfield  {author} {\bibinfo {author} {\bibfnamefont {G.~A.}\ \bibnamefont
  {Zvyagina}}, \bibinfo {author} {\bibfnamefont {T.~N.}\ \bibnamefont
  {Gaydamak}}, \bibinfo {author} {\bibfnamefont {K.~R.}\ \bibnamefont
  {Zhekov}}, \bibinfo {author} {\bibfnamefont {I.~V.}\ \bibnamefont {Bilich}},
  \bibinfo {author} {\bibfnamefont {V.~D.}\ \bibnamefont {Fil}}, \bibinfo
  {author} {\bibfnamefont {D.~A.}\ \bibnamefont {Chareev}}, \ and\ \bibinfo
  {author} {\bibfnamefont {A.~N.}\ \bibnamefont {Vasiliev}},\ }\href@noop {}
  {\bibfield  {journal} {\bibinfo  {journal} {European Physics Letters}\
  }\textbf {\bibinfo {volume} {101}},\ \bibinfo {pages} {56005} (\bibinfo
  {year} {2013})}\BibitemShut {NoStop}%
\bibitem [{\citenamefont {Hosoi}\ \emph {et~al.}(2016)\citenamefont {Hosoi},
  \citenamefont {Matsuura}, \citenamefont {Ishida}, \citenamefont {Wang},
  \citenamefont {Mizukami}, \citenamefont {Watashige}, \citenamefont
  {Kasahara}, \citenamefont {Matsuda},\ and\ \citenamefont
  {Shibauchi}}]{Hosoi16}%
  \BibitemOpen
  \bibfield  {author} {\bibinfo {author} {\bibfnamefont {S.}~\bibnamefont
  {Hosoi}}, \bibinfo {author} {\bibfnamefont {K.}~\bibnamefont {Matsuura}},
  \bibinfo {author} {\bibfnamefont {K.}~\bibnamefont {Ishida}}, \bibinfo
  {author} {\bibfnamefont {H.}~\bibnamefont {Wang}}, \bibinfo {author}
  {\bibfnamefont {Y.}~\bibnamefont {Mizukami}}, \bibinfo {author}
  {\bibfnamefont {T.}~\bibnamefont {Watashige}}, \bibinfo {author}
  {\bibfnamefont {S.}~\bibnamefont {Kasahara}}, \bibinfo {author}
  {\bibfnamefont {Y.}~\bibnamefont {Matsuda}}, \ and\ \bibinfo {author}
  {\bibfnamefont {T.}~\bibnamefont {Shibauchi}},\ }\href@noop {} {\bibfield
  {journal} {\bibinfo  {journal} {Proceedings of the National Academy of
  Sciences}\ }\textbf {\bibinfo {volume} {113}},\ \bibinfo {pages} {8139}
  (\bibinfo {year} {2016})}\BibitemShut {NoStop}%
\bibitem [{\citenamefont {Watson}\ \emph {et~al.}(2015)\citenamefont {Watson},
  \citenamefont {Kim}, \citenamefont {Haghighirad}, \citenamefont {Davies},
  \citenamefont {McCollam}, \citenamefont {Narayanan}, \citenamefont {Blake},
  \citenamefont {Chen}, \citenamefont {Ghannadzadeh}, \citenamefont
  {Schofield}, \citenamefont {Hoesch}, \citenamefont {Meingast}, \citenamefont
  {Wolf},\ and\ \citenamefont {Coldea}}]{Watson15}%
  \BibitemOpen
  \bibfield  {author} {\bibinfo {author} {\bibfnamefont {M.~D.}\ \bibnamefont
  {Watson}}, \bibinfo {author} {\bibfnamefont {T.~K.}\ \bibnamefont {Kim}},
  \bibinfo {author} {\bibfnamefont {A.~A.}\ \bibnamefont {Haghighirad}},
  \bibinfo {author} {\bibfnamefont {N.~R.}\ \bibnamefont {Davies}}, \bibinfo
  {author} {\bibfnamefont {A.}~\bibnamefont {McCollam}}, \bibinfo {author}
  {\bibfnamefont {A.}~\bibnamefont {Narayanan}}, \bibinfo {author}
  {\bibfnamefont {S.~F.}\ \bibnamefont {Blake}}, \bibinfo {author}
  {\bibfnamefont {Y.~L.}\ \bibnamefont {Chen}}, \bibinfo {author}
  {\bibfnamefont {S.}~\bibnamefont {Ghannadzadeh}}, \bibinfo {author}
  {\bibfnamefont {A.~J.}\ \bibnamefont {Schofield}}, \bibinfo {author}
  {\bibfnamefont {M.}~\bibnamefont {Hoesch}}, \bibinfo {author} {\bibfnamefont
  {C.}~\bibnamefont {Meingast}}, \bibinfo {author} {\bibfnamefont
  {T.}~\bibnamefont {Wolf}}, \ and\ \bibinfo {author} {\bibfnamefont {A.~I.}\
  \bibnamefont {Coldea}},\ }\href@noop {} {\bibfield  {journal} {\bibinfo
  {journal} {Physical Review B}\ }\textbf {\bibinfo {volume} {91}},\ \bibinfo
  {pages} {155106} (\bibinfo {year} {2015})}\BibitemShut {NoStop}%
\bibitem [{\citenamefont {Park}\ \emph {et~al.}(2018)\citenamefont {Park},
  \citenamefont {Sakai}, \citenamefont {Erten}, \citenamefont {Mackenzie},\
  and\ \citenamefont {Hicks}}]{Park18}%
  \BibitemOpen
  \bibfield  {author} {\bibinfo {author} {\bibfnamefont {J.}~\bibnamefont
  {Park}}, \bibinfo {author} {\bibfnamefont {H.}~\bibnamefont {Sakai}},
  \bibinfo {author} {\bibfnamefont {O.}~\bibnamefont {Erten}}, \bibinfo
  {author} {\bibfnamefont {A.~P.}\ \bibnamefont {Mackenzie}}, \ and\ \bibinfo
  {author} {\bibfnamefont {C.~W.}\ \bibnamefont {Hicks}},\ }\href {\doibase
  10.1103/PhysRevB.97.024411} {\bibfield  {journal} {\bibinfo  {journal}
  {Physical Review B}\ }\textbf {\bibinfo {volume} {97}},\ \bibinfo {pages}
  {024411} (\bibinfo {year} {2018})}\BibitemShut {NoStop}%
\bibitem [{\citenamefont {Zhao}\ \emph {et~al.}(2016)\citenamefont {Zhao},
  \citenamefont {Yelland}, \citenamefont {Bruin}, \citenamefont {Sheiken},
  \citenamefont {Canfield}, \citenamefont {Fritsch}, \citenamefont {Sakai},
  \citenamefont {Mackenzie},\ and\ \citenamefont {Hicks}}]{Zhao16}%
  \BibitemOpen
  \bibfield  {author} {\bibinfo {author} {\bibfnamefont {L.}~\bibnamefont
  {Zhao}}, \bibinfo {author} {\bibfnamefont {E.~A.}\ \bibnamefont {Yelland}},
  \bibinfo {author} {\bibfnamefont {J.~A.}\ \bibnamefont {Bruin}}, \bibinfo
  {author} {\bibfnamefont {I.}~\bibnamefont {Sheiken}}, \bibinfo {author}
  {\bibfnamefont {P.~C.}\ \bibnamefont {Canfield}}, \bibinfo {author}
  {\bibfnamefont {V.}~\bibnamefont {Fritsch}}, \bibinfo {author} {\bibfnamefont
  {H.}~\bibnamefont {Sakai}}, \bibinfo {author} {\bibfnamefont {A.~P.}\
  \bibnamefont {Mackenzie}}, \ and\ \bibinfo {author} {\bibfnamefont {C.~W.}\
  \bibnamefont {Hicks}},\ }\href@noop {} {\bibfield  {journal} {\bibinfo
  {journal} {Physical Review B}\ }\textbf {\bibinfo {volume} {93}},\ \bibinfo
  {pages} {195124} (\bibinfo {year} {2016})}\BibitemShut {NoStop}%
\bibitem [{\citenamefont {Marcus}\ \emph {et~al.}(2018)\citenamefont {Marcus},
  \citenamefont {Kim}, \citenamefont {Tutmaher}, \citenamefont
  {Rodriguez-Rivera}, \citenamefont {Birk}, \citenamefont {Niedermeyer},
  \citenamefont {Lee}, \citenamefont {Fisk},\ and\ \citenamefont
  {Broholm}}]{Marcus18}%
  \BibitemOpen
  \bibfield  {author} {\bibinfo {author} {\bibfnamefont {G.~G.}\ \bibnamefont
  {Marcus}}, \bibinfo {author} {\bibfnamefont {D.-J.}\ \bibnamefont {Kim}},
  \bibinfo {author} {\bibfnamefont {J.~A.}\ \bibnamefont {Tutmaher}}, \bibinfo
  {author} {\bibfnamefont {J.~A.}\ \bibnamefont {Rodriguez-Rivera}}, \bibinfo
  {author} {\bibfnamefont {J.~O.}\ \bibnamefont {Birk}}, \bibinfo {author}
  {\bibfnamefont {C.}~\bibnamefont {Niedermeyer}}, \bibinfo {author}
  {\bibfnamefont {H.}~\bibnamefont {Lee}}, \bibinfo {author} {\bibfnamefont
  {Z.}~\bibnamefont {Fisk}}, \ and\ \bibinfo {author} {\bibfnamefont {C.~L.}\
  \bibnamefont {Broholm}},\ }\href {\doibase 10.1103/PhysRevLett.120.097201}
  {\bibfield  {journal} {\bibinfo  {journal} {Physical Review Letters}\
  }\textbf {\bibinfo {volume} {120}},\ \bibinfo {pages} {097201} (\bibinfo
  {year} {2018})}\BibitemShut {NoStop}%
\end{thebibliography}%

\end{document}